\documentclass{article}

\usepackage{microtype}
\usepackage{graphicx}
\usepackage{booktabs} %
\usepackage{subcaption}

\usepackage[export]{adjustbox}

\usepackage{tikz}
\newcommand*\circled[1]{\raisebox{.5pt}{\textcircled{\raisebox{-1.05pt} {#1}}}}

\usepackage{hyperref}

\usepackage{caption}
\usepackage{hypcap}

\usepackage[preprint]{acl}

\usepackage{times}
\usepackage{latexsym}

\usepackage[T1]{fontenc}

\usepackage{algorithm}
\usepackage{algorithmic}

\usepackage{amsmath}
\usepackage{amssymb}
\usepackage{mathtools}
\usepackage{amsthm}

\usepackage[capitalize,noabbrev]{cleveref}

\usepackage{xspace}

\theoremstyle{plain}

\theoremstyle{definition}

\theoremstyle{remark}

\usepackage[textsize=tiny]{todonotes}
\newcommand{\sori}{T_{\text{o}}}

\newcommand{\sw}{T_{\text{w}}}
\newcommand{\swi}[1]{\sw^{(#1)}}
\newcommand{\ssus}{T_{\text{sus}}}
\newcommand{\ssusi}[1]{T_{\text{sus}}^{(#1)}}
\newcommand{\id}{\mu}
\newcommand{\idspace}{\mathbb{M}}
\newcommand{\idsize}{M}

\newcommand{\metadata}{z}
\newcommand{\algname}{\textsc{Waterfall}\xspace}
\newcommand{\algmbit}{\textsc{M-bit}\xspace}
\newcommand{\algpnlw}{\textsc{P-nlw}\xspace}
\newcommand{\algsrcmk}{\textsc{SrcMarker}\xspace}
\newcommand{\egbasic}{Vec}
\newcommand{\sem}{c}

\newcommand{\textspace}{\mathbb{T}}
\newcommand{\attackspace}{\mathbb{A}}
\newcommand{\vocabspace}{\mathbb{V}}
\newcommand{\permutekspace}{\mathbb{K}_{\pi}}
\newcommand{\permuteksubspace}{K_{\pi}}
\newcommand{\orthkspace}{\mathbb{K}_{p}}
\newcommand{\permutekid}{k_{\pi}}
\newcommand{\orthkid}{k_{p}}
\newcommand{\wmop}{\mathcal{W}}
\newcommand{\vop}{\mathcal{V}}
\newcommand{\semsimilarity}{\mathcal{S}}
\newcommand{\orthop}{\mathcal{F}}
\newcommand{\orthfn}{\phi}
\newcommand{\pop}{\mathcal{P}}
\newcommand{\aop}{\mathcal{A}}
\newcommand{\avgop}{\bar{\pop}}

\newcommand{\algbasic}{\textsc{Basic}}

\newcommand{\tolsem}{s}

\newcommand{\str}{\kappa}

\newcommand{\llama}{\texttt{llama-2-13b-hf}\xspace}

\def\blankfootnote{\xdef\@thefnmark{}\@footnotetext}

\usepackage{amsmath,amsthm,amsfonts,bm}
\usepackage{IEEEtrantools}

\def\eqref#1{equation~\ref{#1}}

\def\1{\bm{1}}

\DeclareMathAlphabet{\mathsfit}{\encodingdefault}{\sfdefault}{m}{sl}
\SetMathAlphabet{\mathsfit}{bold}{\encodingdefault}{\sfdefault}{bx}{n}

\DeclareMathOperator*{\argmax}{arg\,max}

\theoremstyle{plain}

\theoremstyle{definition}

\usepackage{url}

\usepackage[breaklinks]{hyperref}
\usepackage{cleveref}
\usepackage{graphicx}
\usepackage{multirow}
\usepackage{physics}
\usepackage{longtable}
\usepackage{pifont}

\usepackage{wrapfig}
\usepackage{soul}

\usepackage{amsthm}
\theoremstyle{plain}

\usepackage{listings}
\lstnewenvironment{code}{
    \lstset{
        basicstyle=\small\ttfamily,
        columns=flexible,
        breaklines=true
    }
}{}

\usepackage{rotating}
\usepackage{varwidth}
\usepackage{multicol}
\usepackage{enumitem}
\usepackage{tablefootnote}

\title{
\vspace*{-0.5in}{\normalsize \hfill EMNLP 2024 Main Conference} \\ \vspace{0.2in}
Waterfall: Scalable Framework for Robust Text Watermarking\\ and Provenance for LLMs}

\author{Gregory Kang Ruey Lau\thanks{Equal contribution.}$^{1,2}$, Xinyuan Niu$^{* 1,3}$,
Hieu Dao$^{1}$, Jiangwei Chen$^{1,4}$,\\
\textbf{Chuan-Sheng Foo$^{3,4}$ \quad Bryan Kian Hsiang Low$^{1}$}\\ %
$^1$Department of Computer Science, National University of Singapore\\
$^2$CNRS@CREATE, 1 Create Way, \#08-01 Create Tower, Singapore 138602\\
$^3$Centre for Frontier AI Research;
$^4$Institute for Infocomm Research, A*STAR, Singapore\\
\texttt{\{greglau,niux,daohieu,chenj,lowkh\}@comp.nus.edu.sg},\\
\texttt{foo\_chuan\_sheng@i2r.a-star.edu.sg}\\
}

\begin{document}

\maketitle

\begin{abstract}
    Protecting intellectual property (IP) of text such as articles and code is increasingly important, especially as sophisticated attacks become possible, such as paraphrasing by large language models (LLMs) or even unauthorized training of LLMs on copyrighted text to infringe such IP. However, existing text watermarking methods are not robust enough against such attacks nor scalable to millions of users for practical implementation. In this paper, we propose \algname, the first training-free framework for robust and scalable text watermarking applicable across multiple text types (e.g., articles, code) and languages supportable by LLMs, for general text and LLM data provenance. \algname comprises several key innovations, such as being the first to use LLM as paraphrasers for watermarking along with a novel combination of techniques that are surprisingly effective in achieving robust verifiability and scalability. We empirically demonstrate that \algname achieves significantly better scalability, robust verifiability, and computational efficiency compared to SOTA article-text watermarking methods, and showed how it could be directly applied to the watermarking of code. We also demonstrated that \algname can be used for LLM data provenance, where the watermarks of LLM training data can be detected in LLM output, allowing for detection of unauthorized use of data for LLM training and potentially enabling model-centric watermarking of open-sourced LLMs which has been a limitation of existing LLM watermarking works. Our code is available at \url{https://github.com/aoi3142/Waterfall}.

\end{abstract}
\section{Introduction}

Achieving robust text data provenance via watermarking, independent of its digital format, is an important open problem impacting a wide-ranging set of real-world challenges. Among these is the issue of intellectual property~(IP) enforcement: Content creators of any text format (e.g., articles or code) could potentially combat plagiarism and unauthorized distribution by watermarking their works to prove \textbf{data ownership}. However, existing text watermarking methods have been unable to meet the challenging requirements of many practical problem settings. For example, directly adding digital metadata or invisible Unicode watermarks \citep{rizzoFinegrainWatermarkingIntellectual2019,talebyahvanooeyModernTextHiding2019} may have limited impact in proving text data ownership in adversarial settings as they may be easily removed. Existing natural language watermarking 
\citep{para-nlw, yoo-etal-2023-robust, talebyahvanooeyModernTextHiding2019} that adjusts the text itself to encode IDs are also lack robustness to paraphrasing attacks and have limited scalability in terms of the number of supportable IDs.

Adding to the challenge is the growing prevalence of generative large language models~(LLMs) that may be trained on copyrighted text without permission. To enforce IP rights, content creators need to do \textbf{data provenance for LLMs}, i.e., \emph{prove whether their set of work had been used to train 3rd party black-box LLMs.} 
While there have been recent works tackling these problems \citep{abdelnabiAdversarialWatermarkingTransformer2021, zhang2023remarkllm}, they largely require intervening in the training process of the LLMs. This is unrealistic in practice, as not all LLM service providers may cooperate due to incentive misalignment, and adversaries may use open-source LLMs.

Hence, it is natural to ask \emph{whether it is possible to develop a practical, robust and scalable text watermarking framework 
for protecting IP against both plagiarism and unauthorized training of LLMs}. For example, the watermarks should persist regardless of whether the text has been paraphrased,
converted into speech or handwritten text, or used in unauthorized LLM training (e.g., fine-tuning, in-context learning) to produce a derived output.
The framework should also be general enough to tailor to a wide range of text formats (e.g., natural language or code), and be scalable (i.e., support millions of users with reasonable computational cost).

In this paper, we propose \algname, the first training-free framework for robust and scalable text watermarking applicable across multiple text types (e.g., articles, code) and languages supportable by LLMs, for general text as well as LLM data provenance.
\emph{Rather than viewing LLMs as just sources of IP infringement, we introduce the novel perspective of using LLMs' capabilities to protect existing IP}.    
Though simple, our training-free framework comprises several key innovations such as being the first to use LLM as paraphrasers for watermarking along with a novel combination of techniques that are surprisingly effective in achieving robust verifiability, scalability, and data provenance for LLMs, 
\emph{surpassing state-of-the-art text watermarking methods}. In summary, 
our contributions are as follows:
\begin{enumerate}
    \item We introduced a formulation of the robust and scalable text watermarking problem setting and lay out a set of desiderata to be satisfied (\cref{sec:setting}).
\item To tackle the challenges arising from these desiderata, we proposed \algname comprising novel innovations, including: (a) effective use of LLM paraphrasers to watermark existing text with IP to be protected (\cref{sec:LLMparaphraser}); (b) combination of vocab permutation and a new orthogonal watermarking perturbation method \emph{in token space}, to achieve high scalability and robust verifiability while preserving fidelity (\cref{sec:basis}).

\item We conducted comprehensive empirical evaluations, demonstrating that \algname achieves significantly better scalability, robust verifiability, and computational efficiency compared to SOTA article-text watermarking methods (\cref{sec:exp_article}), while meeting the desiderata for a variety of applications, including for LLM data provenance of articles (\cref{sec:ft}). We also showed how \algname could be directly applied to the watermarking of programming code (\cref{sec:exp_code}).
\end{enumerate}

\section{Problem formulation and Desiderata}\label{sec:setting}

Consider $\idsize$ clients, each with unique watermark ID $\id \in \idspace$ and textual data $\sori \in \textspace$ (e.g., articles or code) represented as token sequences $\sori=[w_1,...,w_N]$, where each token $w_i$ is from an ordered vocab space $\vocabspace=\{v_1,...,v_{|\vocabspace|}\}$.
We assume that $\sori$ has semantic content $\sem$ (e.g., the IP content) that is only determined by its tokens and fully represents the text's value.
Text formatting is irrelevant, especially as adversaries can strip all formatting, making those channels unusable for watermarking\footnote{Attacks include converting text to audio or non-digital formats like written text, which removes format-based watermarks
(e.g., homoglyphs and zero-width Unicode characters) \citep{rizzoFinegrainWatermarkingIntellectual2019} or digital metadata.}.

\noindent \paragraph{Watermarking:} Client $i$ uses a watermarking operator 
$\wmop(\id_i,\sori)\rightarrow \swi{i}$
    to produce a text $\swi{i}$ that contains watermark $\id_i$, preserves $\sem$, and can then be used/distributed freely.

\noindent \paragraph{Attacks:} There are adversaries who aim to claim the IP in $\swi{i}$ through attacks $\aop(\swi{i})\rightarrow \ssusi{i}$ that generate their own text $\ssusi{i}$ without the watermark $\id_i$ while preserving semantic content $\sem$.
Adversaries do not know $\id_i$ but are able to perform several classes of attacks: \\
\noindent
$\attackspace_1$: alter $\swi{i}$ with word addition/removal/substitutions; \\
\noindent
$\attackspace_2$: translate and paraphrase $\swi{i}$ with an LLM; \\
\noindent
$\attackspace_3$: watermark $\swi{i}$ again with a different watermark; \\
\noindent
$\attackspace_4$: using $\swi{i}$ with any LLM for in-context prompting; \\
\noindent
$\attackspace_5$: using $\swi{i}$ to fine-tune any LLM.

\noindent  \paragraph{Verification:} Client $i$ can use a verification operator $\vop(\id_i,\ssus)$ to generate a score $q$ indicating the likelihood that $\ssus$ contains $\mu_i$. They can then use a setting-specific threshold $\bar{q}$ to classify $\ssus$ as watermarked with $\mu_i$ if $q \geq \bar{q}$. The operator $\vop$ should be quick and not assume access to $\sori$, as in practice client $i$ may have a large set of $\sw$ and would need to automate the application of $\vop$ to scan through a large set of $\{\ssus\}$ to identify any plagiarism (further discussion in \cref{app:plagiarism_check}).

Given the above, a suitable watermarking framework should satisfy the following desiderata:

\noindent \paragraph{1. Fidelity.} The watermarked text $\sw$ should be semantically similar to $\sori$, i.e., $\semsimilarity(\sori,\sw)\geq \tolsem$, where $\semsimilarity:\textspace \cross \textspace \rightarrow [0,1]$  is a user-defined fidelity metric depending on the purpose and type of text (e.g., semantic similarity score for articles, or unit tests for code) and $\tolsem$ is a setting-specific threshold. We define $\textspace_{\sem,\tolsem}^{\wmop} = \{T \in \textspace : \semsimilarity(\sori,T) \geq \tolsem \}$ as the support set of all $\sw$ that a watermarking operator $\wmop$ can possibly generate for $\sori$ with content $\sem$ under a $\tolsem$-fidelity setting.

\noindent \paragraph{2. Verifiability.} The verification operator $\vop(\id_i,\swi{i})$ 
should have high efficacy, accounting for Type I and II errors over various thresholds $\bar{q}$. We evaluate this with AUROC computed over a test set.

Note that there is a trade-off between fidelity and verifiability. Applying a stronger, more verifiable watermark tends to 
reduce text fidelity and the optimal setting depends on each use case. We can evaluate a watermarking scheme in general, while taking into account this trade-off, using its fidelity-verifiability Pareto frontier
(e.g., as plotted in \cref{fig:fidelity}a).

\noindent \paragraph{3. Robust verifiability.} The verification operator on watermarked text after attacks $\aop \in \attackspace$, i.e., $\vop(\id_i,\aop(\swi{i}))$, retains high verifiability.
This means that the watermark should remain verifiable even after attacks, which constrains framework design. For example, the verification operator should not extract $\mu$ in any subroutine, as an attacker may use it to get $\mu$ and devise an $\attackspace_3$ attack to overwrite it (see \cref{sec:exp_article}).

\noindent  \paragraph{4. Scalability.} The framework should support a large $|\idspace|$ (set of IDs) while meeting all other desiderata.

\begin{figure}[t]
    \centering
    \includegraphics[width = \columnwidth, ]{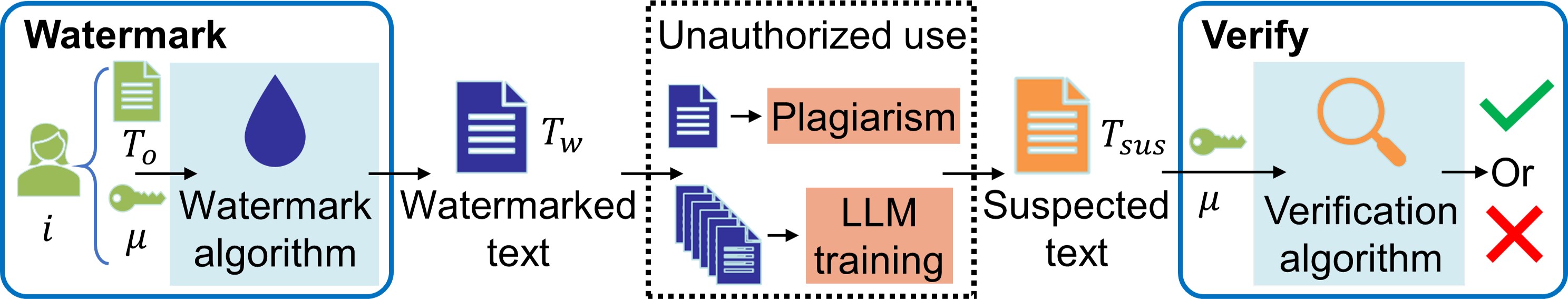}
\vskip -0.05in
    \caption{
    Schematics of problem formulation.
    Client $i$ watermark text $\sori$ with ID $\id_i$
    to watermarked text $\swi{i}$. After manipulation by a third party, client can verify watermark in $\ssus$.
    }\label{fig:formulation}
\vskip -0.06in
\end{figure}

\section{Method}\label{sec:method}

We discuss three key insights to tackle challenges arising from these desiderata, before combining these to present our framework 
\algname (\underline{Water}marking \underline{F}ramework \underline{A}pplying \underline{L}arge \underline{L}anguage Models).

\subsection{Increasing support set for watermarking via LLMs} \label{sec:LLMparaphraser}

First, note that the \textbf{fidelity} desideratum is a major constraint to a scheme's ability to meet the other desiderata. Intuitively, a scheme that can only generate a small set $\textspace_{\sem,\tolsem}^{\wmop}$ of possible watermarked text would have fewer ways to encode the watermark, leading to lower signal capacity (smaller $|\idspace|$, lower \textbf{scalability}), and less capacity for error correction to withstand attacks (lower \textbf{robust verifiability}).

For illustration, consider a basic semantic watermarking scheme (\algbasic) that lists out synonyms for each word in the original text $\sori$ (e.g., big cat) and remembers a map of IDs
to possible combinations of these synonyms (e.g., 01:big feline, 10:large cat, 11:large feline). Watermarking for ID $\id$ is then selecting the text $\sw$ with the matching synonym combination. 
Note that schemes like \algbasic~typically only have a relatively small support set $\textspace_{\sem,\tolsem}^{\wmop}$
and hence limited watermarking possibilities.

However, LLMs can come up with many more possibilities and access a larger $\textspace_{\sem,\tolsem}^{\wmop}$ compared to schemes like \algbasic~using mechanical paraphrasing rules (e.g., synonym replacement).
Past works have shown that LLMs
can effectively paraphrase text given suitable prompts \citep{shu2023rewritelm, witteveen-andrews-2019-paraphrasing}. For example, while synonym replacement can only generate possibilities involving word replacements, an LLM may be able to completely reorder,  break, or fuse sentences while preserving semantic content $\sem$.
In general, as some expressions are more common, we can associate a probability distribution $p_{\sem}(T)$ over this set $\textspace_{\sem,\tolsem}^{\wmop}$.

Intuitively, we can consider a suitable paraphrasing prompt combined with text $\sori$ as tokens $\hat{\sem}$ that can constrain an LLM's text generation to $\textspace_{\sem,\tolsem}^{\wmop}$.
Given $\hat{\sem}$, the LLM autoregressively access $p_{\sem}(T)$ by producing conditional probability distributions $p(w_j|\hat{w}_{1:j-1},\hat{\sem})$ for token $w_j$ at step $j$ given the preceding sampled tokens $\hat{w}$, and sampling for each step until it deemed that it had conveyed $\sem$.
Specifically, at step $j$, the LLM generates a vector of logits $L_j(\hat{w}_{1:j-1},\hat{\sem}) 
\in
\mathbb{R}^{|\vocabspace|}$, where
\begin{equation} \label{eq:LLMpara}
    p(w_j|\hat{w}_{1:j-1},\hat{\sem}) = \text{softmax}(L_j(\hat{w}_{1:j-1},\hat{\sem})).
\end{equation}
We denote LLMs used this way as \emph{LLM paraphrasers}. 
By using LLM paraphrasers,
we 
significantly increase 
$\textspace_{\sem,\tolsem}^{\wmop}$, which helps us better meet the fidelity, robust verifiability and scalability desiderata. 

\begin{figure*}[ht]
\centering
\includegraphics[width=\linewidth]{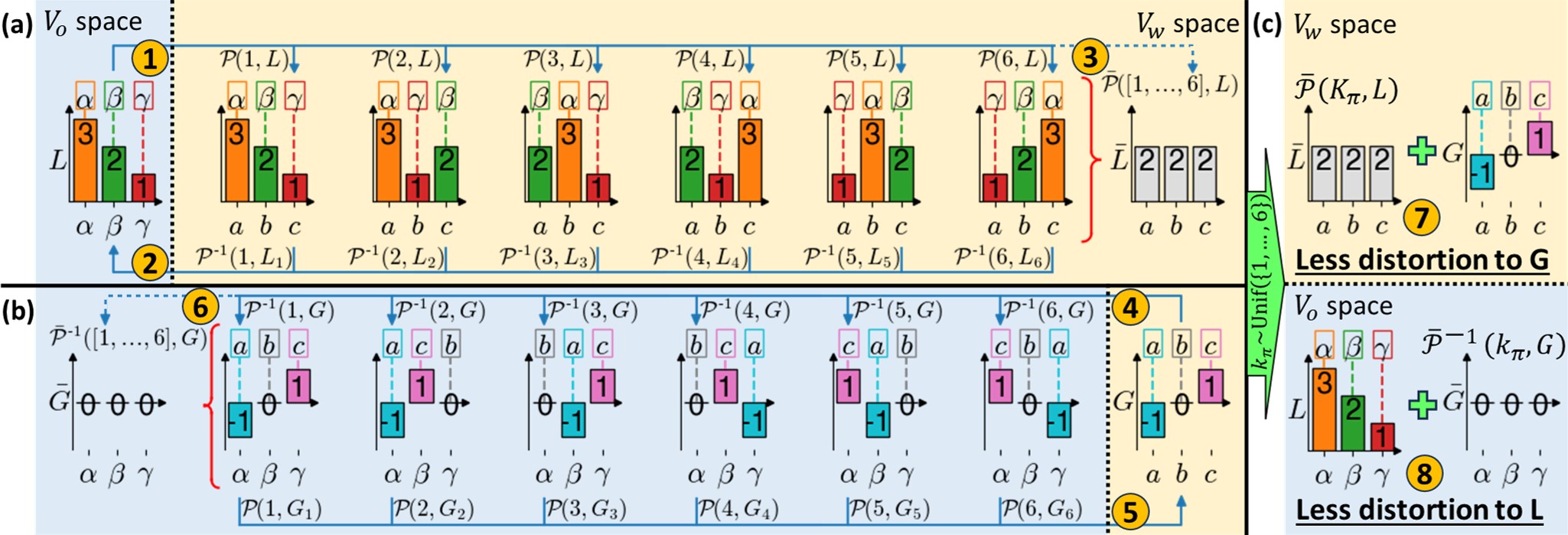}
\caption{
Intuition on permutation operators $\pop$, $\pop^{-1}$ applied on LLM logits $L$ and watermarking signal $G$ with toy example, \egbasic.
\textbf{(a)} $\pop$ applied to $L$ in the $V_o$ space results in 6 possible permutations in $V_w$ space. This averages to constant vector $\bar{L}$.
\textbf{(b)} Similarly, $\pop^{-1}$ applied to $G$ in $V_w$ produces permutations in $V_o$. These averages to constant vector $\bar{G}$.
\textbf{(c)} With $\permutekid$ sampled uniformly from the possible keys $K_\pi$ over multiple LLM generation steps, $L+G$ in shows less distortion to $G$ in $V_w$ space, and to $L$ in $V_o$ space.
}
\label{fig:permutation}
\end{figure*}

\subsection{Increasing robustness using \texorpdfstring{$n$}--gram watermarking with LLM deviation correction} \label{sec:method_n_gram}

Given the extensive threat model, most watermarking schemes would face a major challenge in meeting the \textbf{robust verifiability} desideratum. For example, $\attackspace_2$ paraphrasing attacks would likely break schemes such as \algbasic~which depend on word ordering\footnote{Using example in \cref{sec:LLMparaphraser}, ``large cat"$\rightarrow$``cat that is large" would invert the embedded ID ``10" to ``01".}, 
let alone attacks involving further processing by black-box LLMs (e.g., $\attackspace_4$, $\attackspace_5$ attacks).
Instead, we could decompose $p_{\sem}(T)$ and the watermarked text $\sw$ into multiple signal carriers, and embed the same watermarking signal to all. This way, we adopt a probabilistic approach where each carrier \emph{could independently be used to verify a watermark}, to withstand attacks that can only corrupt a proportion of carriers.

Specifically, we could consider each consecutive $n$ tokens in $\sw$ as an $n$-gram carrier unit. At each LLM paraphraser token generation step $j$, we could apply a 
watermarking operator $\wmop$ (\cref{sec:basis}) that 
perturbs the logits of \cref{eq:LLMpara} 
based on the ID $\id$ and past $n-1$ generated tokens: $\check{L}_j=\wmop[\id,\hat{w}_{j-n+1:j-1}](L_j(\hat{w}_{1:j-1},\hat{\sem}))$.
The perturbed logits will cause a detectable bias in each $n$-gram,
hence the more $n$-grams that persist after any attack, the higher the verifiability. 

Meanwhile, in future generation steps $j'$, the \emph{LLM paraphraser will correct deviations from semantic content $\sem$ and preserve fidelity} given sufficient generation steps, as the subsequent logits $L_{j'}(\hat{w}_{1:{j'}-1},\hat{\sem})$ are still conditioned on paraphrasing prompt $\hat{\sem}$.

This approach increases our framework's robustness against not just paraphrasing attacks, but also more general LLM-based attacks (e.g., $\attackspace_5$).
Past works have shown that language models tend to generate few novel $n$-grams outside their training set for small $n$ \citep{mccoyHowMuchLanguage2021}. Hence, LLMs trained on text with our watermarked $n$-grams may more likely generate them in their output. 
Given sufficient queries to these LLMs, the watermark could then be reliably verified, which we empirically demonstrate in \cref{sec:exp}.

\subsection{Increasing scalability with vocab permutation and orthogonal perturbation} \label{sec:basis}

Finally, we propose a watermarking operator $\wmop$ comprising two components:
1) vocab permutation, and 2) orthogonal perturbation. 
In this section, we will use a toy example (\egbasic) to show how these components work before presenting their general form. In \egbasic, we have logits $L=[3,2,1]$, indexed by an ordered set $V_o=\{\alpha,\beta, \gamma\}$ representing the token space, e.g., $L(\alpha)$ = 3. 
\cref{fig:permutation}a presents $L$ as a graph ($V_o$ as $x$-axis).

\paragraph{Vocab permutation. } 
The vocab permutation operator $\pop$ produces a single permutation of $V_o$ and $L$ for any given key $\permutekid$ (arrow \circled{1} in \cref{fig:permutation}a). The inverse operator $\pop^{-1}$ reverses the permutation of $\pop$ when provided the same key (arrow \circled{2} in \cref{fig:permutation}a). As $|V_o|=3$, there are 6 possible permutations
of $L$, plotted as graphs over a new ordered index $V_w=\{a,b,c\}$, which we can interpret as the watermarking space. Then, we define the average permutation operator $\avgop$ acting on $L$ (indexed by $V_o$) as one that takes a sequence of keys $K_\pi$, apply $\pop$ to get $L_{\permutekid}$ for each $\permutekid \in K_\pi$, and averages them to get a vector $\bar{L}$ (indexed by $V_w$). Note that when we use $\avgop$ on $L$ over all possible keys, we get a constant vector (e.g., $\bar{L}=\sum_{i=1}^{6} L_i/6 = [2,2,2]$, \circled{3} in \cref{fig:permutation}a).

Similarly, given a vector $G$ indexed by $V_w$, which we can interpret as the watermark signal, the inverse operator $\pop^{-1}$ permutes $G$ and $V_w$ given a key $\permutekid$, mapping it to $V_o$, the LLM-ordered token space (arrow \circled{4} in \cref{fig:permutation}b). $\avgop^{-1}$ 
acting on $G$ analogously averages over all keys, and will also give a constant vector indexed over $V_o$ (e.g., $\bar{G}=\sum_{i=1}^{6} G_i/6 = [0,0,0]$, \circled{6} in \cref{fig:permutation}b).

This leads to an interesting insight: the permutation operators provide a way for us to \emph{add watermark signals to logits in 
a deterministically shifting $V_w$ space (based on a sequence of keys) to boost verifiability and fidelity}.
For illustration, assume that an LLM paraphraser produces $L$ (in $V_o$-space) for all token generation steps. We use a long sequence $K_\pi$ of pseudo-random uniformly sampled keys to apply $\pop$ on $L$ multiple times ($n$-gram watermarking), and add the same watermarking signal $G$ in each resulting $V_w$ space for all instances. If we apply $\avgop^{-1}$ with $K_\pi$ on the perturbed signal $L+G$, the distortion from the permuted $L$ will effectively contribute only uniform background noise to $G$ (\circled{7} in \cref{fig:permutation}c), which improves \textbf{verifiability}. If we instead convert $L+G$ back to $V_o$ space (for token generation) with $\pop^{-1}$ for all steps and apply $\avgop$, we get the original logits with only uniform background noise from watermarking (\circled{8} in \cref{fig:permutation}c), which improves \textbf{fidelity}.

\begin{figure*}[ht]
    \centering
    \begin{subcaptiongroup}
    \parbox[b]{.49\textwidth}{
        \centering
        \includegraphics[width=\linewidth]{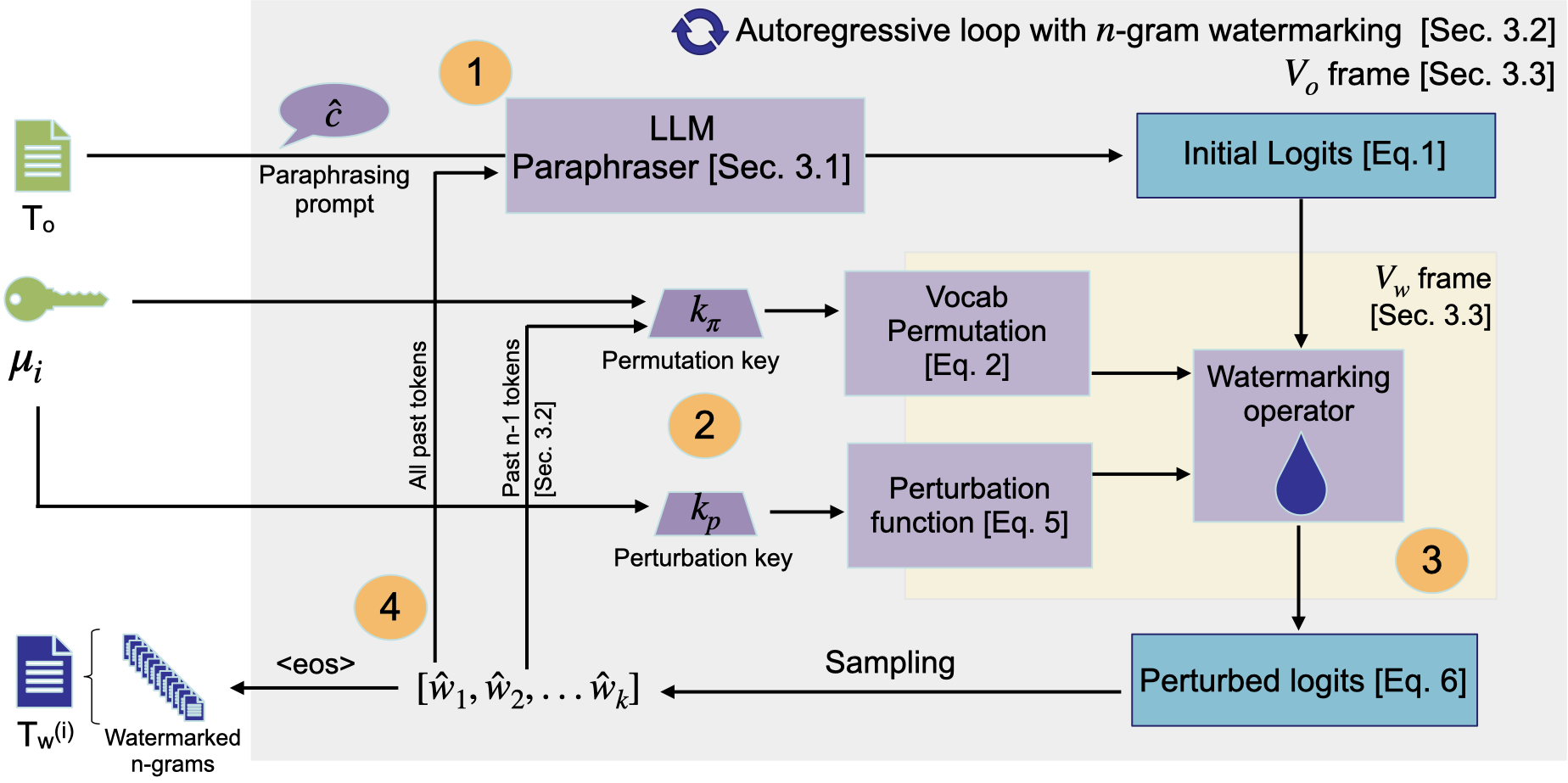}
    }
    \hfill
    \parbox[b]{.5\textwidth}{
        \centering
        \includegraphics[width=\linewidth]{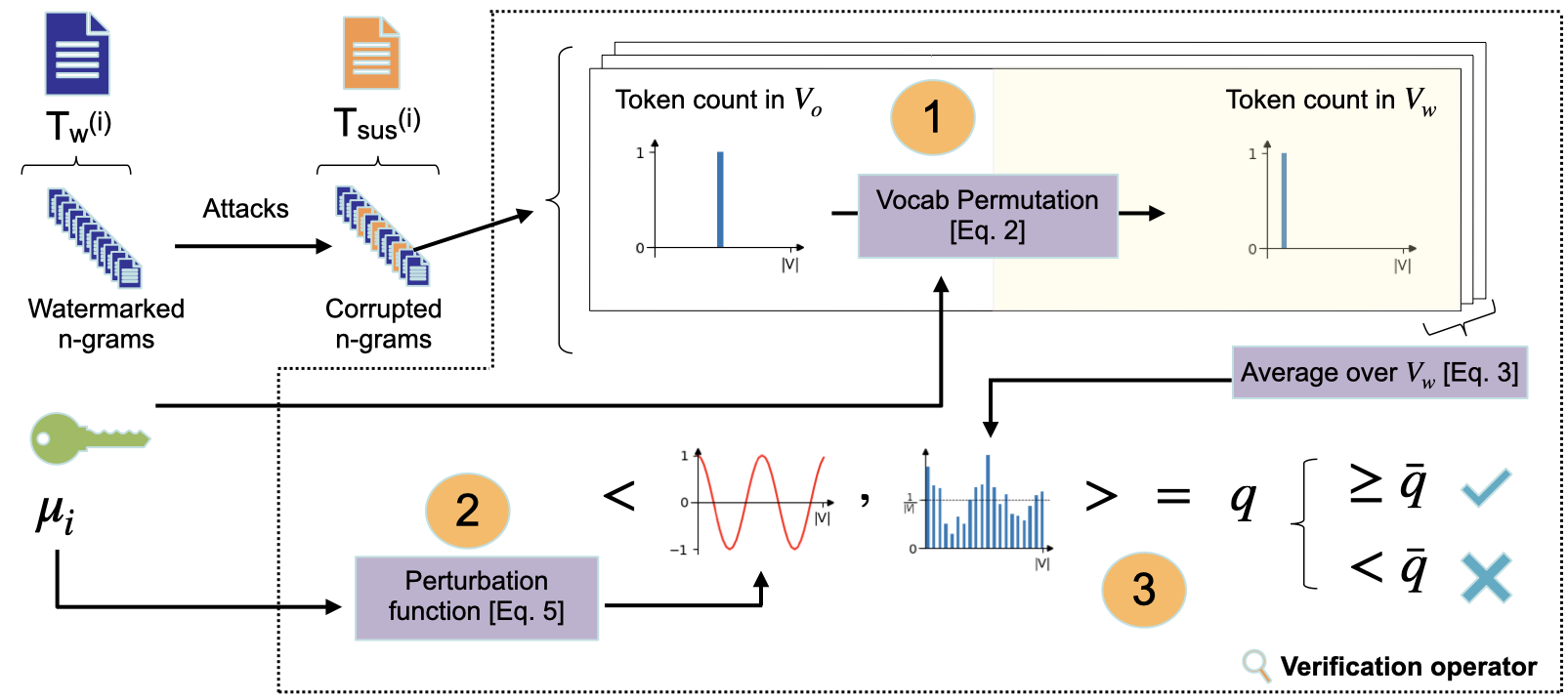}
    }
    \end{subcaptiongroup}
    \caption{\textbf{Left}: Watermarking schematic. \circled{1} LLM paraphraser takes in $T_o$, produces initial logits. \circled{2} $\permutekid$ and $\orthkid$ from ID $\mu$ and metadata $\orthkid$ for vocab permutation and perturbation function. \circled{3} Perturb logits with \cref{eq:perturbed_logits}. \circled{4} Sample perturbed logits, feed past tokens to the next iteration. \textbf{Right}: Verification schematic. \circled{1} Permute tokens from $\ssus$ into $V_w$ with $\mu$ and preceding $n-1$ tokens, to get average cumulative distribution. \circled{2} Compute perturbation function $\orthop_1(k_p)$ linked to $\mu$. \circled{3} Compute verification score as inner product of $\orthop_1(k_p)$ and cumulative distribution, and compare with threshold.}
    \label{fig:framework}
\end{figure*}

More generally, we define the vocab permutation operator $\pop$
and its inverse $\pop^{-1}$ as pseudorandom permutations over ordered sets $V_o$ and $V_w$ given a key $\permutekid \in \permutekspace$:
\begin{align}
    \label{eq:permute} 
    \pop(\permutekid,V_o) &= V_o^{\permutekid} \nonumber \\ 
    \pop^{-1}(\permutekid,V_w) &= V_w^{\permutekid} \nonumber \\
    \pop^{-1}(\permutekid,\pop(\permutekid,V_o))&=V_o,
\end{align}
where $V_o^{\permutekid}$, $V_w^{\permutekid}$ are uniform-randomly chosen permutations of $V_o$ and $V_w$ if $\permutekid$ is sampled randomly.
For a function $L$ over $V_o$ mapped to a vector of length $|V_o|$, we have 
$L(\pop(\permutekid,V_o))=L(V_o^{\permutekid})$ 
and we overload notation by defining
$\pop(\permutekid,L(\cdot))\triangleq L(\pop(\permutekid,\cdot))=L_{\permutekid}$.
As in the \egbasic~example, $\pop$ applied to a function (vector) can be viewed as the same function but with its domain permuted. 

We then define an average operator $\avgop$ over a sequence of keys $\permuteksubspace$ acting on a function $L$,
\begin{equation}
    \avgop(\permuteksubspace,L) \triangleq {\textstyle\frac{1}{|\permuteksubspace|}\sum_{k_{\pi}\in \permuteksubspace}} \pop(\permutekid,L),
\end{equation}
which outputs an average function of $L$ over $V_w$ (denoted as $\bar{L}$ ). 
$\avgop(\permuteksubspace,L)$ will flatten towards a constant function over $V_w$ for a sufficiently large $\permuteksubspace$.
To achieve 
this
for our framework, we set $\permuteksubspace=\{ \permutekid\mid\permutekid=h_{\pi}(\mu, \hat{w}_{j-n+1:j-1})\}_j$,
for all LLM paraphrasing steps $j$
and where $h_{\pi}$ is a hash function, which generates pseudorandom $K_\pi$ sequences. 
Empirically, we clearly observe the flattened and clear watermarking signals (see \cref{fig:empirical_illustration_T_o} in Appendix).

\paragraph{Orthogonal perturbation:} Our proposed perturbation operator $\orthop$ involves two sub-operations acting on $V_w$. It first maps each key $\orthkid \in \orthkspace$ to a unique function in a pre-defined family of orthogonal functions, and then adds the chosen perturbation function to the logits $L_j$ of the LLM output in $V_w$ space:
\begin{align}
    \orthop_1: \orthkspace \hookrightarrow \{\orthfn : V_w \rightarrow \mathbb{R}^{|V_w|} \mid \langle \phi_i,\phi_l\rangle = \delta_{il}\} \\
    \label{eq:perturb_func}
    \orthop(\orthkid,\str,L_j) = L_j+\str \orthop_1(\orthkid)
\end{align}
where $\langle \cdot,\cdot \rangle$ denotes the canonical dot product over $V_w$.
Examples of orthogonal function families include the Fourier or square wave basis, discretized over $\vocabspace$. The key $\orthkid=h_{p}(\mu,\metadata) \in \orthkspace$ is a client defined function $h_p$ of ID $\id$, and also any metadata $\metadata$ (which could be extracted after verification as we demonstrate in \cref{sec:exp_article}) if required. $\str$ is a scalar that controls the perturbation magnitude.

Combining both components, our watermarking operator (\cref{fig:framework}, and \cref{alg:watermark} in Appendix) for generation step $j$ involves (a) using $\permutekid=h_{\pi}(\mu, \hat{w}_{i-n+1:i-1})$ and the permutation operator $\pop(\permutekid,L_j)$ to transform logits from the $V_o$ to $V_w$ space, (b) applying the perturbation operator in \cref{eq:perturb_func}, and (c) transforming the perturbed logits back to $V_o$ space using $\pop^{-1}(\permutekid,.)$ to produce a probability distribution for sampling and generation of the watermarked text $\sw$: 
\begin{align} \label{eq:perturbed_logits}
    \check{L}_j &= \wmop(\permutekid,\orthkid,L_j) \notag \\
    &= \pop^{-1} (\permutekid,\orthop(\orthkid,\str,\pop(\permutekid,L_j))). 
\end{align}

Our verification operator will produce a score by computing the average cumulative token distribution of a text using $\avgop(\permuteksubspace,.)$
and taking the inner product with $\orthop_1(\orthkid)$. Applying the right keys $\orthkid$ and $\permutekid$ on the suspected text $\ssus$ will result in a high score $q$, 
else the score will be close to 0 
(see \cref{fig:framework}, and \cref{alg:verify} in Appendix). 
Using orthogonal functions helps us improve verifiability by avoiding interference from other watermarks (e.g., 
added by adversaries as an $\attackspace_3$ attack). 

Notice that the many possible vocab permutations ($|\vocabspace|!$) and perturbation functions in any orthogonal function family $|\orthop_1|$ allows for a much large set of IDs compared to schemes like \algbasic,
helping with \textbf{scalability}. For example, up to $|\orthop_1|\cdot|\vocabspace|!$ IDs can be assigned to a unique permutation-perturbation function pair for watermarking. Using a relatively small $|\vocabspace|=32000$ and the Fourier basis over that would yield a maximum $|\idspace|\sim 10^{130274}$. Schemes like \algbasic~only support $\idsize$ that scales with the number of possible synonym replacements for a given text.

In addition, with orthogonal functions,
our framework also allows for the embedding of metadata during watermarking. For example, a client can use $\mu$ to verify that $\ssus$ is watermarked, and also extract information on which article it was plagiarized from (\cref{alg:extract}). We demonstrate this empirically in \cref{sec:exp_article} using the Fourier basis as perturbation functions and Discrete Fourier Transform (DFT) for extraction.

\subsection{\algname Framework}\label{sec:framework}

Our watermarking framework, \algname, combines these insights into a structured watermarking/verification process. 
For watermarking (\cref{fig:framework} left), given $\sori$ and $\mu$, \algname uses an LLM paraphraser to autoregressively paraphrase a text $\sori$, producing initial logits for the new text $\sw$ [Step \circled{1}].
The ID $\mu$ is used to seed the vocab permutation operator (\cref{eq:permute}) for mapping the logits to $V_w$ space, and chooses the perturbation function 
(\cref{eq:perturb_func}) [Step \circled{2}], 
both of which will be used in the watermarking operation (\cref{eq:perturbed_logits}) to produce the perturbed logits [Step \circled{3}]. 
The LLM samples the perturbed logits to produce a watermarked token, and for the next token loop, the past $n-1$ tokens are used to seed vocab permutation while all past tokens are fed as context which helps the LLM paraphraser maintain $\sw$ fidelity despite watermarking [Step \circled{4}]. 

For verification (\cref{fig:framework} right), each token in $\ssus$ is counted in $V_w$ space as specified by $\mu$ and the previous tokens in the same $n$-gram unit, producing an average cumulative token distribution [Step \circled{1}]. The ID $\mu$ also
specifies a specific perturbation function [Step \circled{2}], which is used to perform an inner product with the cumulative distribution to compute a verification score $q$ [Step \circled{3}].

\textbf{Practical considerations. } \algname is highly adaptable, i.e., it can be implemented with different LLM as paraphrasers, allowing our framework to achieve better watermarking performance and support more text types as the LLM landscape evolves.
Methods like prompt engineering \citep{wei2023chainofthought,lin2023instinct}
and Reflexion \citep{reflexion,self-refine} may also help to boost performance in some settings, as we demonstrate in our code watermarking experiments (\cref{app:code_rf}).
Additional engineering techniques applied to the LLM paraphraser generation process could also further improve the performance of \algname. For example, using beam search \citep{shao2017generating} instead of just token-level sampling and choosing the best watermarked text version that balances between fidelity and verifiability results in significantly better watermarking, beyond the results presented in \cref{sec:exp}. We elaborate further on this implementation, along with possible large-scale deployment methods of \algname and other practical considerations in \cref{app:practical_deployment}.

\section{Experiments}\label{sec:exp}

\begin{figure}[t]
\centering
\resizebox{\columnwidth}{!}{
\centering\includegraphics{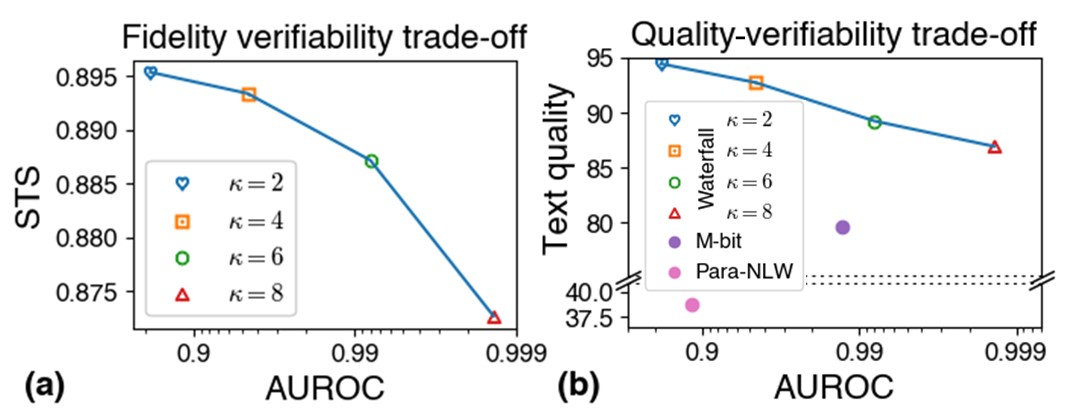}
}
\caption{Higher watermarking strength $\str$ improves verifiability and extraction accuracy. 
(a) Increasing $\str$ trades off fidelity for higher verifiability. (b) \algname performs significantly better than benchmarks in text quality, achieving a text quality-verifiabilty Pareto frontier that is much higher benchmarks (plotted as single points since they do not have adjustable settings to balance the quality-verifiability trade-off).}
\label{fig:fidelity}
\end{figure}

\begin{figure}[t]
\centering
\resizebox{\columnwidth}{!}{
\centering\includegraphics{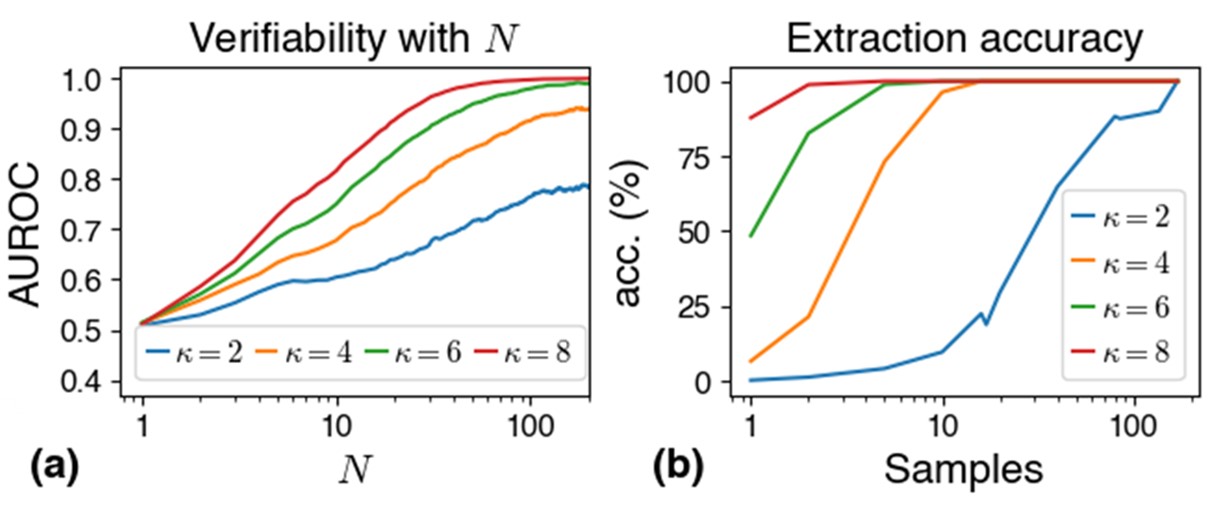}
}
\caption{(a) Longer token length $N$ improves verifiability. (b) Combining more pieces of text improves extraction accuracy towards $100\%$. Extraction accuracy is significantly higher than random guess accuracy of $0.003125\%$.}
\label{fig:ver_extract}
\end{figure}

\subsection{Data ownership}\label{sec:exp_article}

For watermarking of text articles, we demonstrate the effectiveness of \algname with experiments using text samples $\sori$ from the \texttt{c4 realnewslike} dataset \citep{2019t5}, comprising articles with mean token length of 412. The experiments mirror realistic scenarios, for e.g., news outlets watermarking their articles before publishing them to be able to effectively scan the internet for, and verify, plagiarized content \citep{brewster2023plagiarism}. For this setting, we evaluate the semantic similarity $\semsimilarity$ using the Semantic Textual Similarity (STS) score based on the \texttt{all-mpnet-base-v2} model\footnote{https://huggingface.co/sentence-transformers/all-mpnet-base-v2} ($\semsimilarity$ for sample text pairs are provided in \cref{sec:STS_comparison}).

For benchmarks, we consider two recent linguistics-based watermarking methods: \algmbit by \citet{yoo-etal-2023-robust} and \algpnlw by \citet{para-nlw}. These methods are advanced variants of \algbasic~that use deep learning to improve watermarking performance (details in \cref{sec:alt_wop}).
To implement \algname, we use \llama\footnote{https://huggingface.co/meta-llama/Llama-2-7b-chat-hf} as the paraphraser, and the Fourier basis for the perturbation functions. Additional details such as paraphrasing prompts are in \cref{sec:exp_details}.

\paragraph{Fidelity-verifiability. } We consider the fidelity and verifiability of the schemes before adversarial attacks. Verifiability is computed as the AUROC based on varying their respective classification thresholds, i.e., the verification score threshold $\bar{q}$ for \algname, and bit-error rate threshold for \algmbit and \algpnlw.

\algname allows for adjustable watermarking strength to calibrate the fidelity-verifiability trade-off based on the clients' needs. \cref{fig:fidelity}a shows the Pareto frontier of the trade-off. Stronger watermark strength $\str$ improves verifiability but also introduces larger distortions to the LLM paraphrasing process, decreasing the fidelity of watermarked text.
For our experiments, we mainly used $\str=6$, achieving a mean AUROC of 0.992 and STS of 0.887. Even for shorter texts of just 100 tokens ($about$ 65 words), \algname achieves high verifiability with an AUROC of 0.98 (\cref{fig:fidelity}b). Additional results are in \cref{sec:verifiability_appendix}.
Note that \algmbit and \algpnlw 
allows for only a single fidelity-verifiability score, with mean STS scores of 0.998 and 0.942 respectively, and corresponding AUROC scores of 0.987 and 0.882.
While the STS scores are high, it is expected as the schemes only make minor edits to $\sori$ which would be more fragile to attacks, as we will see later.

\paragraph{Text fluency and quality.} In fact, both \algmbit and \algpnlw causes significant degradation in text fluency and quality despite their high STS scores, significantly impacting its practicality in real-life applications. For example, the word replacements by \algmbit and \algpnlw introduced noticeable linguistic errors that are not captured by the STS score (shown in \cref{sec:STS_comparison}). In contrast, \algname consistently produce high quality watermarked text. 
To systematically analyze this, we quantify text quality via evaluations using ChatGPT, which has been shown to be a good evaluation metric for text fluency \citep{wang2023chatgpt}. Details of the evaluation method, along with results using another text quality metric (GPTScore \citep{fu2024gptscore}) which further corroborate our findings, are in \cref{app:text_quality}. 
We can see in \cref{fig:fidelity}b that \algname significantly outperforms benchmark methods for the fluency score across the different watermark strengths while achieving high verifiability (AUROC), and the results of M-bit and P-nlw lie far within the Pareto frontier of \algname.

\paragraph{Robust verifiability. } We consider the various classes of attacks $\attackspace$ mentioned in \cref{sec:setting}. Details of the setup for each attack and additional results are in \cref{app:attacks}.

\begin{figure}[t]
\centering
\resizebox{\columnwidth}{!}{
\centering\includegraphics{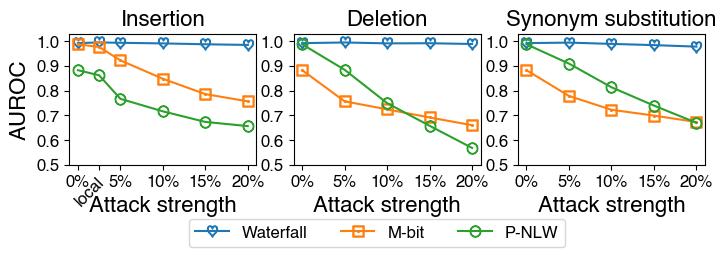}
}
\caption{
\algname demonstrates robust verifiability under $\attackspace_1$ (insertion, deletion, and synonym substitution attacks) with minimal degradation in AUROC compared to \algmbit and \algpnlw.
}
\label{fig:insertion}
\end{figure}

$\attackspace_1$ attacks are insertion, deletion, and synonym substitution attacks that are often considered in past works
As shown in \cref{fig:insertion}, robust verifiability of \algname shows only a slight decrease even with strong attacks on $20\%$ of words, while that of \algmbit and \algpnlw fall drastically with increasing attack strength.

$\attackspace_2$ involves translation and paraphrasing attacks, which are more realistic and effective attacks that can achieve higher 
fidelity
and verification reduction than $\attackspace_1$ and had not been considered by past text watermarking works. We perform \textbf{translation} attack
to translate the watermarked text to Spanish and back to English, and \textbf{paraphrasing} attack
to paraphrase the watermarked text. Again, the verifiability of \algname remains significantly higher than benchmarks post-attack.

$\attackspace_3$ involves using the same scheme to try overwrite the existing watermark with another watermark. For \algname, the 1\textsuperscript{st} watermark remain verifiable even after the 2\textsuperscript{nd} is added, given the design of $\pop$ and $\orthop$ with vocab permutation and orthogonal perturbation functions
that minimizes interference of the 2\textsuperscript{nd} watermark on the 1\textsuperscript{st}.
However, this attack destroys the verifiability of \algmbit and \algpnlw, as the 2\textsuperscript{nd} watermark process almost always chooses the same word positions as the original process, overwriting $\id_1$. 
Furthermore, the benchmark schemes extracts $\id_1$ as part of verification, enabling targeted overlap watermark attacks which we demonstrated in \cref{sec:complement_attack}.

$\attackspace_4$ uses $\sw$ for in-context prompting of any LLM to perform tasks that rely on the IP or semantic content of $\sw$.
For illustration, we considered the case where adversaries use an LLM
to answer questions regarding watermarked articles. As this attack totally changed the structure of the texts, the watermarks of \algmbit and \algpnlw were removed.
However, with \algname, watermarks were still verifiable due to the preservation of watermarked $n$-grams from the context to the response.

$\attackspace_5$ which involves using text containing IP for unauthorized LLM training such as fine-tuning will be discussed in \cref{sec:ft}.

\begin{table}
    \centering
    \caption{
    Robust verifiability under attacks: 
    translation $\attackspace_{2-T}$, paraphrase $\attackspace_{2-P}$, overlap watermark $\attackspace_3$ and in-context prompting $\attackspace_4$.
    }
\begin{small}
\setlength\tabcolsep{3.5pt}
    \begin{tabular}{lcccccccc}
\toprule
         & Pre-attack & $\attackspace_{2-T}$ & $\attackspace_{2-P}$& $\attackspace_3$& $\attackspace_4$\\
\midrule
         \algname& \textbf{0.992} & \textbf{0.951} & \textbf{0.881} & \textbf{0.815} & \textbf{0.775}\\
         \algpnlw& 0.885 & 0.475& 0.508& 0.724 & 0.502\\
         \algmbit& 0.988 & 0.567& 0.363& 0.664 & 0.525\\
\bottomrule
    \end{tabular}
\end{small}
    \label{tab:advanced_attack}
\end{table}

\paragraph{Scalability.}
As mentioned in \cref{sec:basis}, \algname has a large maximum scalability
of $M=|\orthop_1|\cdot|\vocabspace|!\sim 10^{130274}$ based on our implementation using the Fourier perturbation function, and Llama-2 model as LLM paraphraser. 
In comparison, \algmbit and \algpnlw, have scalability dependent on the number of possible synonym replacements in any given text, which is limited by text length and varies for different text.
On the \texttt{c4} dataset with a mean article length of 355 words, \algmbit and \algpnlw can only embed a mean of 9.5 bits ($M \sim 10^{3}$) and 23.2 bits ($M \sim 10^{10}$) respectively.

In practice, scalability is further limited by how well the schemes can differentiate among similar watermarks. For e.g., the verification operation of \algmbit~and \algpnlw~may not be able to distinguish 2 IDs differing by 1 bit (see \cref{app:benchmark_scalability} for details). To demonstrate this, we watermarked $\swi{i}$ with $\mu_i$ and computed the verifiability of $\swi{i}$ against 1000 randomly selected IDs $\id_{j \neq i}$. We found that for \algname, all of the IDs achieved very high AUROC, while \algmbit and \algpnlw have many IDs with low AUROC: 
The 1\textsuperscript{st} percentile AUROC for \algname, \algmbit, \algpnlw are 0.976, 0.614, 0.766 respectively.
Details and further results on scalability up to 100,000 IDs are in \cref{sec:scalability_appendix}.

\paragraph{Metadata extraction.} We also demonstrate how \algname could be used to embed metadata while watermarking. We consider metadata $k_p \in \{1,2,...,31999\}$, and the task is to extract the embedded $k_p$ if the text has been verified as watermarked with $\mu$. We do so by using $k_p$ as the frequency of the Fourier perturbation function $\orthop_1$, and perform extraction with the DFT. \cref{fig:ver_extract}b shows the extraction accuracy of \algname for different perturbation magnitudes $\str$. By taking multiple samples of $\sw$, multiple articles could be combined together to improve extraction accuracy. For $\str=6$, accuracy increases from 48\% to 99\% with only 5 pieces of text. Details are in \cref{app:extraction}.

\paragraph{Computational costs.} We note that \algname also has lower computational cost compared to benchmarks (\cref{tab:comptime}). 
\algname verification can be run in parallel on a CPU, requiring only 0.035s when ran on a single 16-core CPU, which is 75x and 4237x faster than \algmbit and \algpnlw respectively, both which require inference using deep learning models. This is 
important in the context of protection of IP, e.g.,
where data providers may have to scan through large amount of online data for any IP infringement.
Further discussion on the deployment costs are in \cref{app:practical_deployment}.

\begin{table}
\vskip -.01in
\caption{Mean compute time over 100 texts on 1 Nvidia RTX A5000. *Note that verification for \algname was performed only on CPU without requiring a GPU.}
    \centering
\vskip -.02in
\begin{small}
    \begin{tabular}{cccc}
    \toprule
         &  \algname&  \algmbit& \algpnlw\\
         \midrule
         Watermark&  24.8s&  \textbf{2.97s}& 147s\\
         Verification& \textbf{0.035s*}&  2.61s& 148s\\
         \bottomrule
    \end{tabular}
    \label{tab:comptime}
\end{small}
\vskip -.01in
\end{table}

\subsection{Watermarking of code}\label{sec:exp_code}

\begin{table}
\setlength\tabcolsep{1.975pt}
\caption{Fidelity, Verifiability, and Robust Verifiability of \algname with $\str=3$ on code watermarking. 
    }
    \centering
\resizebox{\columnwidth}{!}{%
        \begin{tabular}{lcccc}
\toprule
             &\multirow{2}{*}{\begin{tabular}{@{}c@{}}Fidelity \\ (Pass@10)\end{tabular}} & \multicolumn{2}{c}{Verifiability (AUROC)} &\multirow{2}{*}{\begin{tabular}[c]{@{}c@{}}Scalability\\ (\# of users)\end{tabular}} \\ \cline{3-4} 
             &                        & \multicolumn{1}{l}{Pre-attack} & Post-attack \\
\midrule

            \algsrcmk & 0.984 & 0.726 & 0.662 & $10^5$\\
             \algname & 0.969 & 0.904 & 0.718 & $10^{130274}$ \\
\bottomrule
        \end{tabular}
    \label{tab:code_table}
}
\end{table}

To demonstrate the versatility of \algname, we also consider its out-of-the-box performance on code watermarking.
We used the MBJSP dataset \citep{mbxp_athiwaratkun2022} 
 , and evaluate fidelity $\semsimilarity(\sori,\sw)$ using the pass@10 metric \citep{passk_metric,chen2021codex} achieved by $\sw$ on functional tests for the original code $\sori$. We compare \algname, implemented using 
\texttt{Phind-CodeLlama-34B-v2}\footnote{https://huggingface.co/Phind/Phind-CodeLlama-34B-v2}
as the paraphraser, with \algsrcmk \citep{srcmarker}, a recent state-of-the-art code watermarking scheme, configured for 16-bit watermarks. Experimental details are in \cref{app:code_app}.

We found that surprisingly, \algname achieves higher verifiability and robust verifiability (after $\attackspace_2$ LLM paraphrasing attacks) compared to \algsrcmk while maintaining high code fidelity (\cref{tab:code_table}). This is despite \algname not requiring any manual training/engineering of programming language-specific watermarking rules, which \algsrcmk does. Instead, \algname inherits its code capabilities from its LLM paraphraser, making it easily adaptable to other languages (e.g., see \cref{app:code_python} for Python code results).

\subsection{LLM data provenance of articles} \label{sec:ft}
\begin{figure}[t]
    \centering
    \includegraphics[width=0.8\linewidth]{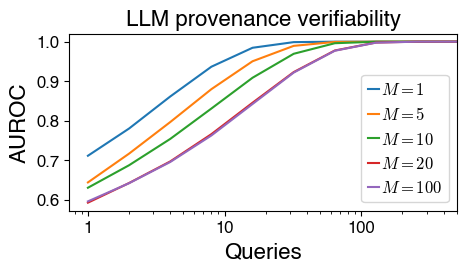}
    \caption{More queries improve LLM attribution verifiability. Increasing number of clients $M$ in the training dataset only slightly decreases verifiability. Note that the curve for $M=20$ overlap with that of $M=100$.}
    \label{fig:attack_5}
\end{figure}
Finally, we explore how \algname watermarks may persist after LLM fine-tuning, allowing us to use them for LLM data provenance. We consider the setting where client $i$ watermarks a set of text $\{\swi{i}\}$ that adversaries use, without authorization, to fine-tune their own LLMs (i.e., $\attackspace_5$ attacks). 
Given multiple queries to the fine-tuned black-box LLM, the goal is for client $i$ to be able to verify that $\{\swi{i}\}$ had been used for training.
This setting mirrors realistic scenarios where content owners want to detect unauthorized use of data for LLM training \citep{Novet_2024}.

For our experiments, we watermarked the ArXiv dataset \citep{clement2019arxiv} which consists of scientific paper abstracts categorized into topics. Each topic category is associated with a unique client ID $\mu$ with $4000$ text. These texts are then used to fine-tune \texttt{gpt2-xl}\footnote{https://huggingface.co/openai-community/gpt2-xl} using the LoRA framework \citep{hu2022lora}\footnote{Note that this is a different model compared to that used for watermarking. We chose this to demonstrate that our watermark can persist despite the models' different tokenizers.} (details in \cref{sec:lora_setup}). We verified that using the watermarked dataset instead of the original dataset has minimal effect on the fidelity of the fine-tuned model (details in \cref{sec:attribution_fidelity}).

\paragraph{Verifiability. } To evaluate verifiability, we queried the fine-tuned model with the first 50 tokens of a randomly chosen abstract, and applied the verifiability operator on the next 100 generated new tokens to test for the associated watermark (details in \cref{sec:attribution_verifiability}). Our results, presented in \cref{fig:attack_5}, shows that \algname has high verifiability, reaching AUROC of 1.0 with just 100 queries to the fine-tuned LLM.

\paragraph{Scalability. }
To explore the scalability of \algname for data provenance, we combined the datasets of different number of clients, $M \in \{1,5,10,20,100\}$, each watermarked with their own unique ID $\id$, and use the combined dataset for fine-tuning the adversarial model. As expected, \cref{fig:attack_5} shows that dealing with an aggregated dataset mixed with a larger $M$ number of different watermarks would result in a decrease in verifiability. However, our results indicate that this decrease leveled off from $M=20$ to $M=100$ and still allow for an AUROC (verifiability) of 1.0 with around 100 queries even for $M=100$, demonstrating the scalability of \algname to a sizable number of clients.

\section{Related Work}\label{sec:related_works}

Early text watermarking techniques \citep{kamaruddinReviewTextWatermarking2018, talebyahvanooeyModernTextHiding2019} primarily depend on structural adjustments (e.g., text formatting, use of different Unicode characters \citep{rizzoFinegrainWatermarkingIntellectual2019}), image-based techniques (e.g., pixel-adjustments of text), or 
semantic watermarking (e.g., substituting synonyms like \algbasic~described in \cref{sec:LLMparaphraser}). 
Recent works have augmented the latter with deep learning and language models for better performance \citep{para-nlw,yoo-etal-2023-robust,ueokaFrustratinglyEasyEditbased2021, abdelnabiAdversarialWatermarkingTransformer2021}. However, as we showed in our experiments, these schemes are not robust to the range of practical LLM-enabled attacks possible today.

A recently popular but separate line of work has focused on the different \emph{model-centric} problem setting of watermarking newly-generated output generated by a single LLM
\citep{kirchenbauerWatermarkLargeLanguage2023,venugopalWatermarkingOutputsStructured2011,christUndetectableWatermarksLanguage2023, kuditipudiRobustDistortionfreeWatermarks2023,zhaoProtectingLanguageGeneration2023}, rather than 
existing text owned by many clients. 
Hence, these works do not address our problem desiderata such as achieving scalability and robust verifiability while requiring semantic preservation of the original text.
Our work focused on 
\emph{data-centric text watermarking} of original text
is the first to use LLM paraphrasers with a novel combination of techniques that are surprisingly effective in addressing the text data ownership and LLM data provenance settings.
For further elaboration on the differences, see \cref{app:comparison_model_centric}.

\section{Discussion and Conclusion}

We proposed \algname, the first training-free framework for text watermarking that has low computational cost, scalability to large number of clients, and robustness to LLM attacks including unauthorized training of LLMs that generates IP-infringing text.

There is currently a lack of actual, practical large-scale deployment of text watermarking effective against LLM attacks, given the current SOTA watermarking methods' limitations and resource requirements. However, \algname may possibly provide a foundation for achieving large-scale deployment, with both decentralized or centralized options. This is made achievable given \algname's low computational cost, scalability to a large number of clients, and robustness to LLM attacks including unauthorized training of LLMs that generates IP-infringing text. 

Our framework highlights a few perspectives that we hope more
would consider. First, \emph{while increasingly capable LLMs allows for easier and more sophisticated forms of potential IP infringement, LLMs themselves could also enable better text IP protection of original texts}. A key strength of \algname is that its capabilities grow as LLMs become more powerful, with increasingly better watermarking performance, allowing it to potentially keep up with the increasing capabilities adversaries can use for IP infringement. It is able to achieve a higher  fidelity-verifiability Pareto frontier, and reduce any fidelity degradation while using higher watermarking strength for greater robust verifiability.

Second, as open-source LLM models become more prevalent and capable, adversaries could directly use them for IP attacks rather than depend on the services of closed-source LLM providers, allowing them to bypass any IP protection measures that these providers may implement \citep{Piper_2024}.
As such, \emph{content creators cannot just rely on LLM providers to assist in IP protection, but instead be equipped with methods such as} \algname \emph{to protect their work before dissemination}, such as by injecting robust watermarks that allows verifiability even after both traditional attacks and unauthorized use in LLM training by adversaries.

Third, a general text watermarking framework like \algname that can apply across different text types and languages not only helps with practical deployment, but also makes it highly versatile and not dependent on any text-specific properties. This makes it \emph{easily adaptable for incorporating new defense methods, providing a strong foundation for future works to build on as new threats emerge}.

\section{Limitations}
As \algname relies on the adjustment of the original text to add watermarks, it may not applicable to all types of text. For example, \algname faces limitations in its application to works where their IP values lie in their style or format (e.g., poems), unless additional methods are applied that cause LLMs to largely preserve such styles while paraphrasing these works, such as optimizing for better paraphrasing prompts to be used with more capable LLMs, or iteratively refining the text through multiple rounds of watermarking. We further discuss examples of such methods in \cref{app:practical_deployment}, and also demonstrate an example of using the Reflexion technique \cite{reflexion} to overcome similar issues for code in \cref{app:code_rf}. 

Similar to other linguistics-based text watermarking methods, \algname would also not be applicable where changes to the text are unacceptable (e.g. lyrics of a country's national anthem), or when applied to very short text (e.g. messages of just a few tokens). Nevertheless, \algname is still useful for a wide range of settings where the IP lies mainly in the content of the text, and presents a step forward for practical deployment of text watermarking. Future work could build on \algname to adapt it to other use cases for data provenance, such as data currency (i.e., ensuring that the data is up-to-date) or data authenticity (i.e., that the data has not been manipulated).

\section*{Acknowledgments}

This research/project is supported by the National Research Foundation, Singapore under its AI Singapore Programme (AISG Award No: AISG2-PhD/2023-01-039J). This research is part of the programme DesCartes and is supported by the National Research Foundation, Prime Minister’s Office, Singapore under its Campus for Research Excellence and Technological Enterprise (CREATE) programme. This research is supported by the National Research Foundation Singapore and the Singapore Ministry of Digital Development and Innovation, National AI Group under the AI Visiting Professorship Programme (award number AIVP-$2024$-$001$).
Xinyuan Niu is supported by the Centre for Frontier AI Research of Agency for Science, Technology and Research (A$^\star$STAR).
Jiangwei Chen is supported by the Institute for Infocomm Research of Agency for Science, Technology and Research (A$^\star$STAR).
We acknowledge CSC (Finland) for awarding this project access to the LUMI supercomputer, owned by the EuroHPC Joint Undertaking, and hosted by CSC (Finland) and the LUMI consortium. The access was made possible via collaboration between NSCC (Singapore) and CSC (Finland). 
\newpage

\bibliography{LLM}

\newpage
\appendix
\onecolumn

\section{Additional details on watermarking and verification operators}

\begin{algorithm}[htb]
    \caption{\algname Watermarking algorithm}\label{alg:watermark}
    \begin{algorithmic}[1]
    \STATE \textbf{Input:} Original text $\sori$, ID $\id$, text-specific metadata $\metadata$, $n$-gram length $n$, perturbation magnitude $\str$, keys functions $h_{\pi}$ and $h_p$
        \STATE Provide to LLM paraphraser a prompt $\hat{\sem}$ containing $\sori$ and paraphrasing instructions, which represents semantic content $\sem$ of $\sori$.
        \STATE Compute $k_p=h_p(\mu,\metadata)$.
        \FOR {$j = 1,\dots$}
            \STATE Obtain logits $l_j(\hat{w}_{1:j-1},\hat{\sem})$ from LLM paraphraser, given \cref{eq:LLMpara}.
                \STATE Compute $k_{\pi}=h_{\pi}(\id,\hat{w}_{j-n+1:j-1})$.
                \STATE Compute perturbed logits $\check{l_j}$ based on \cref{eq:perturbed_logits}.
                \STATE Sample token $\hat{w_j}$ based on the perturbed probability distribution $\check{p_j}=\text{softmax}(\check{l_j})$.
        \ENDFOR
    \STATE \textbf{Output:} Watermarked text $\sw=[\hat{w}_1,...,\text{\textless eos\textgreater}]$.
    \end{algorithmic}
\end{algorithm} 

\begin{algorithm}[htb]
    \caption{\algname Verification algorithm} \label{alg:verify}
    \begin{algorithmic}[1]
    \STATE \textbf{Input:} Suspected text $\ssus=[\hat{w}_1,\ldots,\hat{w}_N$], ID $\id$, $n$-gram length $n$, keys function $h_{\pi}$, perturbation key $\orthkid$, test threshold $\bar{q}$.
    \STATE Initialize a vector $C$ of length $|V_o|$, which keeps track of token counts, to 0.
        \FOR {$j = 1,\dots, |\ssus|$}
            \STATE Compute $k_{\pi}=h_{\pi}(\id,\hat{w}_{j-n+1:j-1})$ and permutation operator $\pop(k_{\pi})$, given \cref{eq:permute}.
            \STATE Set $C(\pop(k_{\pi},\hat{w}_i))++$.
        \ENDFOR
        \STATE Compute avg cumulative token distribution $\bar{C}=C/N$.
        \STATE Compute verification score $q = \langle \bar{C},\frac{\orthop_1(k_{p})}{\lVert\orthop_1(k_{p})\rVert_2}\rangle$ based on \cref{eq:perturb_func}.
    \STATE \textbf{Output:} Returns true if $q \geq \bar{q}$.
    \end{algorithmic}
\end{algorithm} 

\begin{algorithm}[htb]
    \caption{\algname Extraction algorithm} \label{alg:extract}
    \begin{algorithmic}[1]
    \STATE \textbf{Input:} Suspected text $\ssus=[\hat{w}_1,\ldots,\hat{w}_N$], ID $\id$, $n$-gram length $n$, keys function $h_{\pi}$.
    \STATE Initialize a vector $C$ of length $|V_o|$, which keeps track of token counts, to 0.
        \FOR {$j = 1,\dots, |\ssus|$}
            \STATE Compute $k_{\pi}=h_{\pi}(\id,\hat{w}_{j-n+1:j-1})$ and permutation operator $\pop(k_{\pi})$, given \cref{eq:permute}.
            \STATE Set $C(\pop(k_{\pi},\hat{w}_i))++$.
        \ENDFOR
        \STATE Compute avg cumulative token distribution $\bar{C}=C/N$.
        \STATE Compute highest scoring key $\hat{\orthkid} = \argmax_{\orthkid \in \orthkspace}{\langle \bar{C},\frac{\orthop_1(k_{p})}{\lVert\orthop_1(k_{p})\rVert_2}\rangle}$ based on \cref{eq:perturb_func}.
    \STATE \textbf{Output:} Returns $\hat{\orthkid}$.
    \end{algorithmic}
\end{algorithm}

\section{Empirical illustration of watermarking signal in $\sw$}
Here we empirically illustrate how the watermarking signal can be embedded in $V_w$ space with the background logits appearing as uniform noise, as described in \cref{sec:basis}. To illustrate the presence of the watermarking signal, we use the combined watermarked dataset used in the data ownership experiments, and plot its average cumulative token distribution $\bar{C}$ (in \cref{alg:verify}). 

\begin{figure}[ht]
\centering
\resizebox{0.8\columnwidth}{!}{
\centering\includegraphics{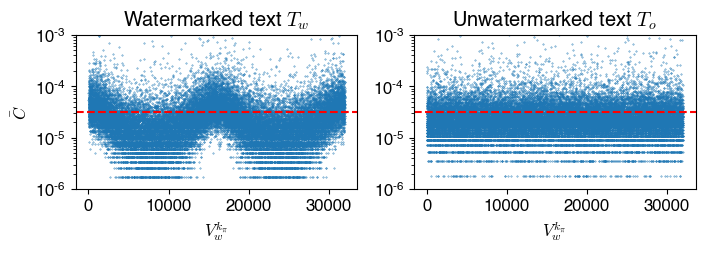}
}
\vskip -0.1in
\caption{Average cumulative token distribution $\bar{C}$ of watermarked and unwatermarked text from subset of \texttt{c4 realnewslike} dataset. Fourier watermark signal with frequency 2 is clearly visible in $\sw$ (left) as compared to $\sori$ (right).}
\label{fig:empirical_illustration_T_o}
\end{figure}

\cref{fig:empirical_illustration_T_o} shows that when we use the correct ID and $\permutekid$ for verification, the watermarking function can be clearly seen for the watermarked text $\sw$ (distribution in the shape of a cosine curve of 2 periods for $\orthkid=2$), while the unwatermarked text $\sori$ shows a flat function. 

\begin{figure}[ht]
\centering
\resizebox{0.8\columnwidth}{!}{
\centering\includegraphics{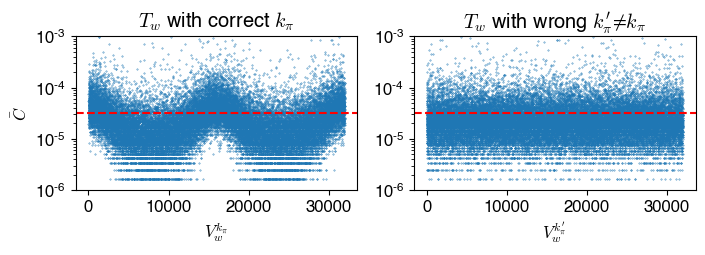}
}
\vskip -0.1in
\caption{Average cumulative token distribution $\bar{C}$ of watermarked text from subset of \texttt{c4 realnewslike} dataset. Fourier watermark signal with frequency 2 is clearly visible when performing the correct permutation $\pop(\permutekid)$ (left) compared to the wrong permutation $\pop(\permutekid')$ (right).}
\label{fig:empirical_illustration_T_w}
\end{figure}

Similarly, \cref{fig:empirical_illustration_T_w} shows that when verifying watermarked text $\sw$, the watermarking function is only visible with the correct permutation $\pop(\permutekid)$ (distribution in the shape of a cosine curve of 2 periods for $\orthkid=2$), but not with a different permutation $\pop(\permutekid')$ (i.e., wrong ID).

\section{Examples of orthogonal watermarking functions} \label{sec:orthgonal_fn_ex}
We chose cosine and sine functions as the watermarking functions, due to the orthogonality between the cosine and sine functions of different frequencies.

\[ \phi_{\orthkid}(j) = \begin{cases}
    \cos\left(2\pi \orthkid \frac{j}{\abs{\vocabspace}}\right) &\text{if } \orthkid \leq \frac{\abs{\vocabspace}}{2} \\
    \sin\left(2\pi (\orthkid-\frac{\abs{\vocabspace}}{2}) \frac{j}{\abs{\vocabspace}}\right) &\text{otherwise} \\
\end{cases} \]

where 
$j \in \{1, \dots, \abs{\vocabspace}\}$ denote the index in the vocab space, 
$\orthkid \in \{1, \dots, \abs{\vocabspace}-1\}$ denote the index of the available orthogonal functions.
We chose the cosine and sine sequences as any other bounded watermarking sequence can be represented by a collection of sinusoidal sequences via the discrete Fourier transform (DFT).

In general, periodic functions of different frequencies could be used as the system of orthogonal functions, along with the phase-shifted counterparts by phase of a quarter wavelength. Other than the cosine and sine functions, one other example is the square wave functions.

Let $k_N = \underset{k \in \mathbb{N}^{*}}{\max} \{ k \mid \abs{\vocabspace} \equiv 0 \ (\text{mod } 2^k)\}$. Assuming $k_N \geq 2$,
the number of orthogonal square waves supported is $2k_N-1$, such that $k_p \in \{1, \dots, 2k_N-1\}$. The square watermarking function is defined as follows.
\[ 
\phi_{\orthkid}(j) = \begin{cases}
    \left(-1\right)^{\lfloor 2^{k_p}\frac{j}{\abs{\vocabspace}} \rfloor} &\text{if }  k_p \leq k_N \\
    \left(-1\right)^{\lfloor 2^{\left(k_p-k_N\right)}\frac{j}{\abs{\vocabspace}} + 0.5 \rfloor} &\text{otherwise} \\
\end{cases} 
\]

\section{Discussion on weaknesses of existing text watermarking methods}\label{app:benchmark_weakness}
Both benchmark text watermarking methods, \algmbit and \algpnlw, are unable to achieve perfect verification performance despite having deterministic watermarking and verification algorithms, as stated in their respective papers, and corroborated in our experiments.

Both methods first use a language model to select viable word positions at which to perform the synonym substitution, then another model or word list to generate the list of possible synonym for substitution. During verification, we observe that the watermark could be corrupted in three ways.

Firstly, as the text being fed to the model for selecting the word replacement location is different (original text during watermarking and watermarked text during verification), the locations being selected during verification could be different as that used for watermarking.

Secondly, even if the correct locations are selected, a different synonym list could be generated during verification, due to the words that were changed at other locations during the watermarking process.

Thirdly, as the benchmarks perform watermarking by sequentially embedding the bits of the watermark ID into the text, any modifications to the text that inserts, deletes or shuffles the text would destroy the watermark ID. If an insertion or deletion error appears early in the text either through the first corruption above or through attacks, i.e., the location for a word replacement being inserted or removed during the verification as compared to during watermarking, the remainder of the watermark ID would be shifted in position, resulting the all the bits after the error to be in the wrong position, resulting in poor verifiability and robust verifiability. Additionally, as illustrated in \cref{sec:method_n_gram}, attacks that reorders the text will also shuffle the watermark ID, destroying its robust verifiability.

On the other hand, \algname is not susceptible to the above mentioned issues. As discussed in \cref{sec:method_n_gram}, the watermark signal is injected into each $n$-gram in the watermarked text, and does not depend on the specific location within the sentence, or specific word replacements. As the hash function $h_\pi$ is deterministic, the same permutation used during watermarking will always be selected during verification, as long as the $n$-gram unit is preserved.

\section{Data ownership experimental setting}\label{sec:exp_details}
\subsection{Dataset}\label{sec:ownership_details}
\begin{figure}[ht]

\centering
\resizebox{0.8\columnwidth}{!}{
\centering\includegraphics{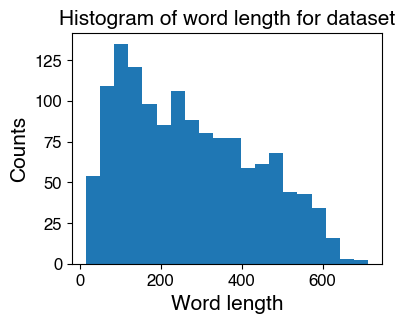}
\centering\includegraphics{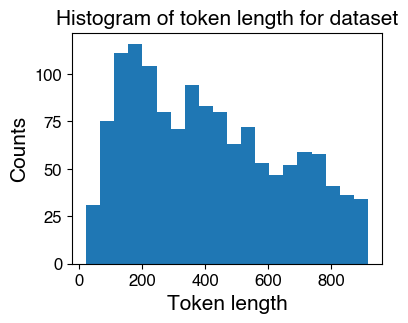}
}
\vskip -0.1in
\caption{Histogram of word and token lengths of text in the \texttt{c4 realnewslike} dataset used for data ownership experiments.}
\label{fig:c4_word_token_len}
\end{figure}

From the first 2000 samples in the c4 dataset, we selected text that were shorter than 1000 tokens long as our text samples $\sori$, totaling 1360 samples. We restricted the token length to ensure the paraphrasing prompt, original text and watermarked text can fit within the context window of the LLM used for paraphrasing. In practice, to overcome this limitation, longer original text could either be first split up into multiple sections to be watermarked, or an LLM with a longer context window could be used.
The distribution of word and token lengths is shown in \cref{fig:c4_word_token_len}.

\subsection{Watermarking methodology}
To perform paraphrasing, we followed the prompt format for \llama, and used the following prompt to perform watermarking. No effort has been made to optimize the prompt. 

\noindent
\begin{minipage}{\textwidth}
\begin{lstlisting}
[INST] <<SYS>>
Paraphrase the user provided text while preserving semantic similarity. Do not include any other sentences in the response, such as explanations of the paraphrasing. Do not summarize.
<</SYS>>

{text} [/INST] 

Here is a paraphrased version of the text while preserving the semantic similarity:
\end{lstlisting}\end{minipage}

For the results in the experimental (\cref{sec:exp_article}), the watermark was performed with ID $\id=0$ and $\orthkid=1$. Results for other $\id$ and $\orthkid$ are reported below.

After watermarking, we perform a simple post-processing step to strip away extraneous generation by the LLM, by filtering out the last sentence or paragraph that contain the following phrases.

\noindent
\begin{minipage}{\linewidth}
\begin{multicols}{4}
\begin{itemize}[leftmargin=*] 
  \item let me know
  \item paraphrase
  \item paraphrasing
  \item other sentences
  \item original text
  \item same information
  \item Note:
  \item Note :
  \item Please note
  \item Please kindly note
  \item Note that I
  \item semantic similar
  \item semantically similar
  \item similar in meaning
  \item Please be aware
  \item the main changes made
  \item Kindly note
  \item Note this does
  \item I have made sure to
\end{itemize}
\end{multicols}
\vskip 0.1 in
\end{minipage}

This list should be customized depending on the content of the text to be watermarked, and LLM used for watermarking. Other methods of cleaning the watermarked text such as prompting the LLM to critic or correct issues within the watermarked text could be employed \citep{reflexion}.

\subsection{Benchmark experiment settings} \label{sec:alt_wop}
\textbf{\algpnlw } \citep{para-nlw} proposes a watermarking process by incorporating a paraphraser-based lexical substitution model. While \textbf{\algmbit} \citep{yoo-etal-2023-robust} carefully chooses the potential original word to replace via finding features that are invariant to minor corruption, and a BERT-based lexical substitution model. We use these two approaches as the benchmark for text watermarking in the data ownership problem setting. 
\paragraph{Key generation}
As default, both \algmbit and \algpnlw use binary keys as watermark signals. The bits for the keys we use for experiments were generated with a seeded pseudo-random number generator. Specifically, we used 0 as the seed to NumPy's Random Generator to generate the key used in the experiments\footnote{NumPy random generator takes in an unsigned int as the seed}.

\subsection{Verifiability} \label{sec:verifiability_appendix}
In this section, given threshold score $\bar{q}$, we define the classification problem as follows. Positive sample: watermarked text $\swi{i}$; Negative sample: unwatermarked text $\sori$; Predictive positive: $\vop \left( \id_i, \ssus \right) \geq \bar{q}$, Predictive negative: $\vop \left( \id_i, \ssus \right) < \bar{q}$.

\begin{figure}[ht]
\resizebox{\columnwidth}{!}{
\centering\includegraphics{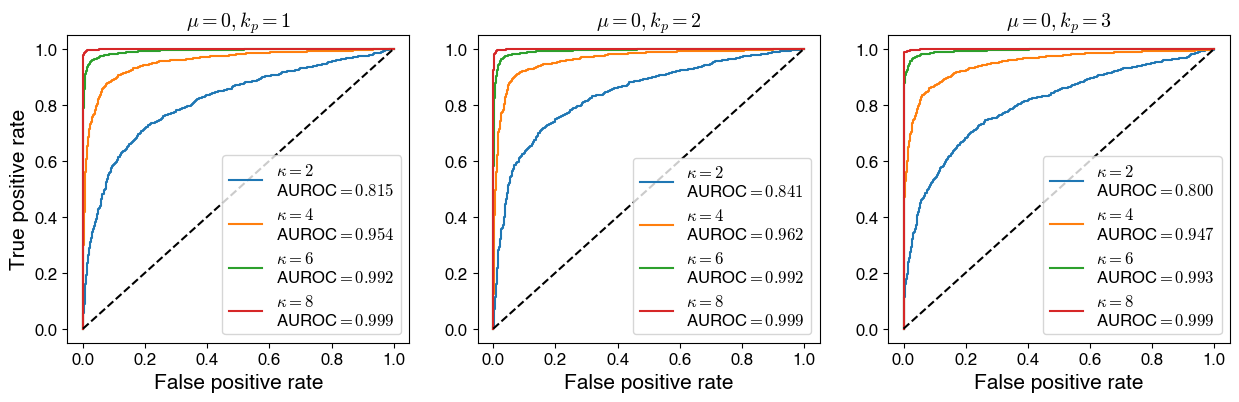}
}
\resizebox{\columnwidth}{!}{
\centering\includegraphics{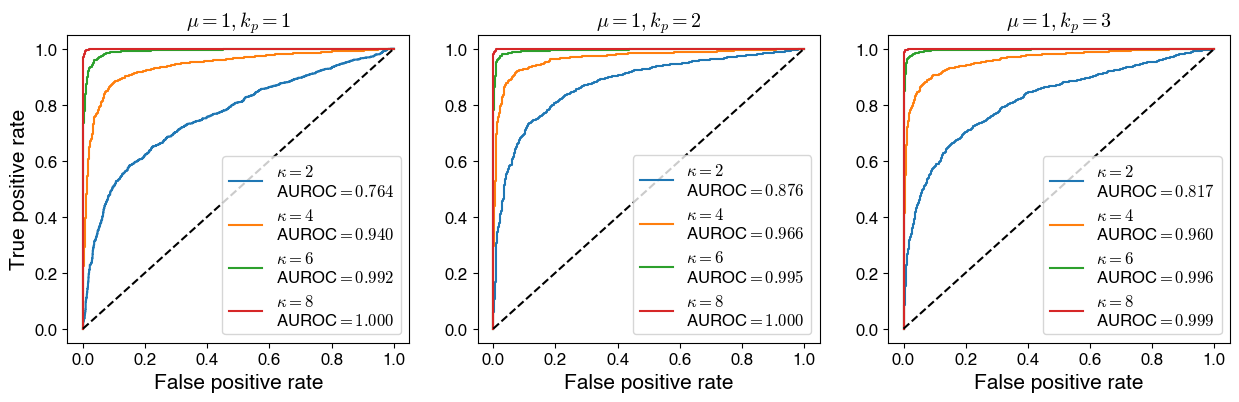}
}
\resizebox{\columnwidth}{!}{
\centering\includegraphics{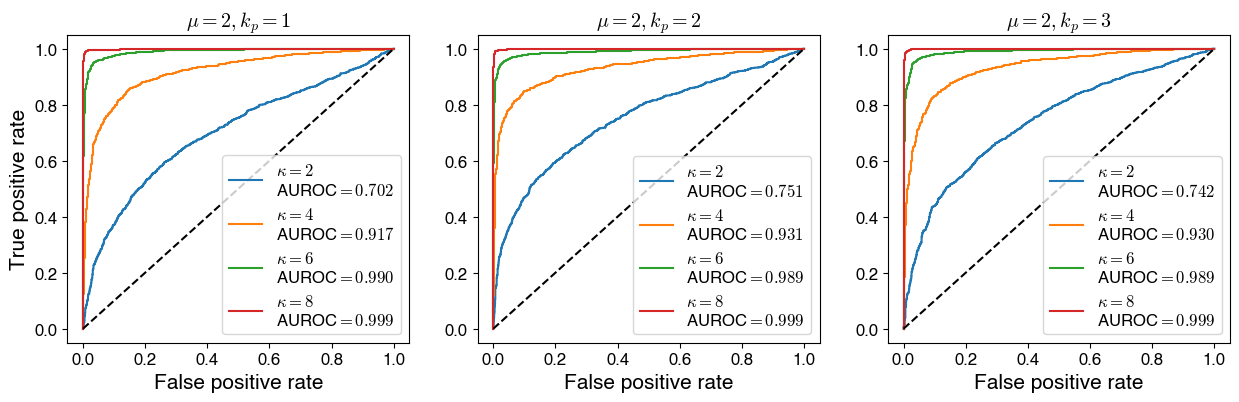}
}
\vskip -0.1in
\caption{ROC curves and corresponding AUROC values for different $\id$, $\orthkid$ and $\str$}
\label{fig:ROC}
\end{figure}

The ROC curves and corresponding AUROC values for different $\id$, $\orthkid$ and $\str$ are shown in \cref{fig:ROC}. We show that verifiability is insensitive to different $\id$, $\orthkid$ used for watermarking. For $\str=6$, \algname was able to achieve AUROC of 0.989-0.996 across the different settings.

\subsection{Fidelity} \label{sec:STS_comparison}
We provide some examples of text watermarked by the \algname, \algmbit and \algpnlw. \cref{tab:c4_samples} shows a few samples from the c4 dataset with watermarked text of varying STS scores. \algmbit has the highest STS across these samples listed, due to its algorithm only changing very few words within the text, resulting in lower scalability as described in \cref{sec:exp_article}. Despite the high STS score, it can be visually seen that text watermarked with \algmbit and \algpnlw introduces linguistic and grammatical errors to the text, which are not measured by the STS score. Further analysis on the text quality is included in \cref{app:text_quality}.

\begin{figure}[H]
\centering
\resizebox{0.4\columnwidth}{!}{
\includegraphics[width=\linewidth]{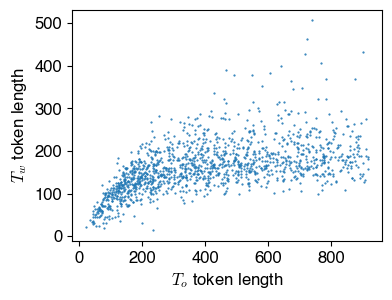}
}
\caption{Distribution of token length of unwatermarked text $\sori$ against watermarked text $\sw$}
\label{fig:un_watermarked_token_length}
\end{figure}

We noticed that there is a tendency of LLMs to summarize when performing paraphrasing, where some details of the text are lost during the watermarking process. This can be seen in the decrease in token length comparing the original unwatermarked text $\sori$ against watermarked text $\sw$ in \cref{fig:un_watermarked_token_length}. However, there are multiple methods of mitigating this issue. Firstly, longer text could be broken apart into different sections to be watermarked separately before being combined together. Secondly, due to the robustness of \algname to modifications, the watermarked text could be edited to correct for errors or reinsert missing information, either manually or by leveraging LLMs to perform self-critic, similar to the method described by \citet{reflexion}. Lastly, as \algname does not require the use or training of specific LLMs for watermarking, it can be applied to newer models with better paraphrasing performance as they become available.

\newpage

\begin{longtable}{|c|p{0.7\columnwidth}|c|}
\caption{Text watermarked with \algname, \algmbit and \algpnlw, along with their STS scores. For \algmbit and \algpnlw, highlighted words are words that were selected for watermarking, and may or may not have been changed as compared to the original text depending on the ID used for watermarking.}
\label{tab:c4_samples}
\\
\hline
         & Text & STS\\
         \hline
\endfirsthead
\hline
         & Text & STS\\
         \hline
\endhead

Original & After the martyrdom of St. Boniface, Vergilius was made Bishop of Salzburg (766 or 767) and laboured successfully for the upbuilding of his diocese as well as for the spread of the Faith in neighbouring heathen countries, especially in Carinthia. He died at Salzburg, 27 November, 789. In 1233 he was canonized by Gregory IX. His doctrine that the earth is a sphere was derived from the teaching of ancient geographers, and his belief in the existence of the antipodes was probably influenced by the accounts which the ancient Irish voyagers gave of their journeys. This, at least, is the opinion of Rettberg ("Kirchengesch. Deutschlands", II, 236). & -\\
\hline
\algname & After St. Boniface's martyrdom, Vergilius became the Bishop of Salzburg in 766 or 767 and worked tirelessly to strengthen his diocese and spread the faith in nearby heathen regions, particularly in Carinthia. He died in Salzburg on November 27, 789. In 1233, he was canonized by Gregory IX. According to Rettberg (``Kirchengesch. Deutschlands'', II, 236), Vergilius' belief that the earth is a sphere was based on the teachings of ancient geographers, and his belief in the existence of the antipodes may have been influenced by the accounts of ancient Irish voyagers. & 0.978\\
\hline
\algmbit (8 bits)& \hl{Following} the martyrdom \hl{of} St. Boniface, Vergilius \hl{became} made Bishop of Salzburg (766 or 767) and laboured successfully for the upbuilding of his diocese as well as for the spread of the Faith in neighbouring heathen countries, especially in Carinthia. He died {at} Salzburg, 27 November, 789. In 1233 he was {canonized} by Gregory IX. His doctrine that the earth is a sphere {was} derived from the teaching of ancient geographers, and his belief in the existence of the antipodes {was} probably influenced by the accounts which the ancient Irish voyagers gave of their journeys. This, at least, is the opinion {of} Rettberg (``Kirchengesch. Deutschlands'', II, 236). & 0.999\\
\hline
\algpnlw (3 bits) & \hl{following} the martyrdom of St. Boniface, Vergilius was made Bishop of Salzburg (766 or 767) and \hl{worked} \hl{worked} for the upbuilding of his diocese as well as for the spread of the Faith in neighbouring heathen countries, especially in Carinthia. He died at Salzburg, 27 November, 789. In 1233 he was canonized by Gregory IX. His doctrine that the earth is a sphere was derived from the teaching of ancient geographers, and his belief in the existence of the antipodes was probably influenced by the accounts which the ancient Irish voyagers gave of their journeys. This, at least, is the opinion of Rettberg (``Kirchengesch. Deutschlands'', II, 236). & 0.964\\
\hline
\pagebreak

Original & Karl Kispert, principal of cyber and information security, has more than 28 years of experience in selling, managing and delivering information risk management, internal audit, regulatory and compliance programs, and information security and technology risk management.
A former chief information security officer, Kispert has helped design and implement cybersecurity programs for many firms, according to the firm.
“By adding this new service line, and bringing someone with Karl’s expertise to the firm, we can service yet another important aspect of our clients’ and prospects’ businesses, ensuring their continued success,” CEO Louis Grassi said in a written statement.
Services will include full security programs, compliance, third party vendor risk assessment, threat management, and managed security services.
& - \\
         \hline
\algname & Karl Kispert, a principal specialist in cybersecurity and information risk management, has extensive experience spanning 28 years in providing sales, management, and delivery of information risk management, internal audit, compliance programs, and technology risk management solutions. As a former Chief Information Security Officer, Kispert has supported the design and implementation of comprehensive cybersecurity programs for numerous organizations. The CEO of the firm, Louis Grassi, has expressed enthusiasm about expanding the firm's service offerings through the integration of this new service line, which will be supported by Kispert's proficiency in providing comprehensive security measures, compliance, vendor risk assessment, threat management, and managed security services.
& 0.899 \\
         \hline
\algmbit (5 bits) & Karl Kispert, principal \hl{in} cyber and information security, has more than 28 years of experience in selling, managing and delivering information risk management, internal audit, regulatory and compliance programs, and information security and technology risk management. A former chief information security officer, Kispert \hl{had} helped design and implement cybersecurity programs for many firms, according to the firm. “\hl{By} adding this new service line, and bringing someone with Karl’s expertise to the firm, we \hl{can} service yet another important aspect of our clients’ and prospects’ businesses, ensuring their continued success,” CEO Louis Grassi said in a written statement. Services \hl{offered} include full security programs, compliance, third party vendor risk assessment, threat management, and managed security services.
& 0.9969 \\
         \hline
\algpnlw (21 bits) & \hl{carl} \hl{kisper}, \hl{principal} of \hl{cyber} and information \hl{protection}, has \hl{has} \hl{than} 28 \hl{old} of experience \hl{experience} \hl{selling}, managing and \hl{delivery} \hl{information} \hl{risk} \hl{risks}, internal audit, \hl{regulatory} \hl{cyber} \hl{cybernetic} programs, and information security and \hl{technology} risk management. A former chief information security officer, Kispert has helped project and \hl{project} cybersecurity programs for many firms, according to the firm. “ By adding this new service line, and bringing someone with Karl’ s expertise to the firm, we can service yet another important aspect of our clients ’ and prospects ’ businesses, ensuring their continued success, \hl{''} CEO Louis Grassi said in a written job. Services will include full security programs, compliance, third party vendor risk assessment, threat management, and \hl{managed} security services.
& 0.938 \\
         \hline
         \pagebreak

Original & Larry checks in with KPCC reporter Sharon McNary, who’s been hitting up several polling stations in Orange County and Los Angeles County, as well as Registrar of Voters for O.C. and L.A.
After being a finalist for LAPD chief in 2009 only to see the job go to Charlie Beck, Michel Moore has been selected to succeed Beck by L.A. Mayor Eric Garcetti.
President Donald Trump signed the “right-to-try” bill into law on Wednesday, a measure that gives terminally ill patients access to experimental drugs that have not yet been approved by the Food and Drug Administration (FDA).
Humans have a habit of measuring things. Our shoe size. The ingredients in our food. How long it takes to get to work, with or without traffic.&-\\
         \hline
\algname & Larry talks with KPCC reporter Sharon McNary about polling stations and the Registrar of Voters in both Orange County and Los Angeles County. The Los Angeles Mayor, Eric Garcetti, has appointed Michel Moore as the new Chief of the LA Police Department after he was previously a finalist for the position in 2009. The US President, Donald Trump, signed a law giving terminally ill patients access to unapproved experimental treatments. Humans tend to quantify aspects of life, such as shoe size, food ingredients, commute times, and more.&0.857\\
         \hline
\algmbit (4 bits) & Larry checks in with KPCC reporter Sharon McNary, who’s been hitting up several polling stations in Orange County and Los Angeles County, as well as Registrar of Voters for O.C. and L.A. After being a finalist for LAPD chief in 2009 only to see the job go to Charlie Beck, Michel Moore has been selected to succeed Beck by L.A. Mayor Eric Garcetti. President Donald Trump signed the “right-to-try” bill into law on Wednesday, a measure that gives terminally ill patients access to experimental drugs that have not yet \hl{become} approved by the Food and Drug Administration (FDA). Humans have a habit \hl{for} measuring things. Our shoe size. The ingredients \hl{of} our food. How long it takes \hl{to} get to work, with or without traffic.&0.999\\
         \hline
\algpnlw (12 bits) & \hl{lary} \hl{controls} \hl{on} \hl{on} KPCC \hl{journalist} Sharon McNary, who \hl{is} s been \hl{attacked} up several polling stations in Orange County and Los Angeles County, \hl{as} well as Registrar of Voters for \hl{O.C}. \hl{los} L.A. After being a finalist for LAPD chief in 2009 only to see the job go to Charlie Beck, Michel Moore has been selected to succeed Beck by L.A. Mayor Eric Garcetti. President Donald Trump \hl{signed} the “ right-to-try ” bill into law on Wednesday, a measure that gives terminally ill patients access to experimental drugs that have not yet been approved by the Food and Drug \hl{on} (FDA). Humans have a habit of measuring things. Our shoe size. The ingredients in our food. How long it takes to get to work, with or without traffic.&0.829\\
         \hline
\pagebreak
Original & Come test your luck on the best slot machine app in the app store. Great graphics make this app so fun to play.
Test your luck with Pharaoh Slots! Bet, Spin and Get Lucky!
& -\\
         \hline
\algname & Experience the ultimate entertainment with the most thrilling slot machine game in the app store! Marvel at stunning visuals that make playing so enjoyable.
& 0.787\\
         \hline
\algmbit (4 bits) & Come test your luck \hl{on} the best slot machine app in the app store. Great graphics make \hl{this} app so fun to play. Test your luck \hl{on} Pharaoh Slots! Bet, Spin \hl{and} Get Lucky!
& 0.9985\\
         \hline
\algpnlw (12 bits) & \hl{please} \hl{test} \hl{yourself} \hl{happiness} \hl{happiness} the best \hl{place} machine app \hl{in} the app store. Great graphics make \hl{it} app \hl{app} fun to play. Test \hl{your} \hl{luck} with Pharaoh Slots ! Bet, Spin and Get \hl{get}!
& 0.716\\
         \hline
\end{longtable}

\subsection{Text quality}
\label{app:text_quality}
To evaluate the fluency and grammatical corectness of the text, we used OpenAI's \texttt{gpt-4o} as an evaluator to judge the text quality of the watermarked text comparing \algname and the benchmarks \algmbit, \algpnlw. Inspired by \citet{wang2023chatgpt}, we prompted \texttt{gpt-4o} with the following prompt.

\noindent
\begin{minipage}{\textwidth}
\begin{lstlisting}
Given the text below, give a score between 0 and 100 for the fluency and grammatical correctness of the text. Be strict and only give high scores sparingly. Only output the number and do not output any other text or explanation.
\end{lstlisting}\end{minipage}

We also evaluated text quality using GPTScore \citep{fu2024gptscore}, which is a metric to evaluate the quality and fluency using the \texttt{GPT3 babbage-002} backbone.

\begin{figure}[ht]
\resizebox{\columnwidth}{!}{
    \includegraphics{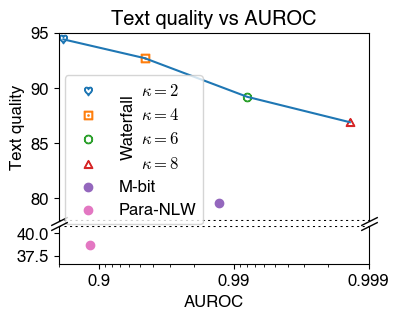}
    \includegraphics{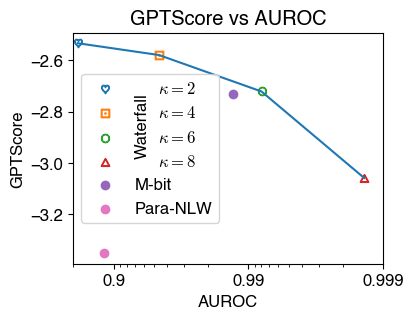}
}
\vskip -0.1in
\caption{Text quality against verifiability trade-off. \textbf{Left}: Fluency score (modified from \citet{wang2023chatgpt}) and \textbf{Right}: GPTScore \citep{fu2024gptscore}}
\label{fig:text_quality}
\end{figure}

In \cref{fig:text_quality}, we show that benchmarks \algmbit and \algpnlw fall well within the Pareto frontier of \algname when considering text quality vs. verifiability trade-off.

\subsection{Verifiability fidelity trade-off} \label{sec:tradeoff_appendix}

\begin{figure}[ht]
\resizebox{\columnwidth}{!}{
    \includegraphics{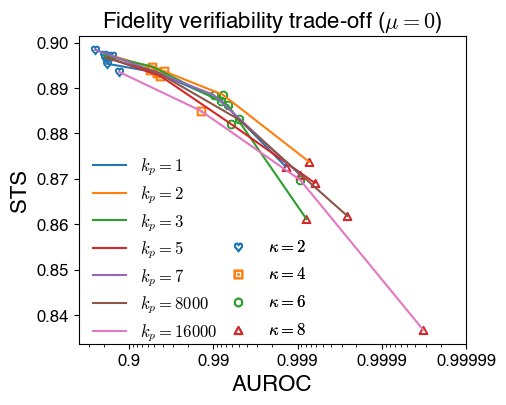}
    \includegraphics{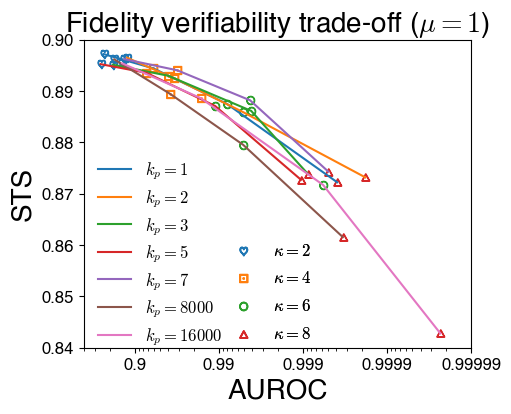}
    \includegraphics{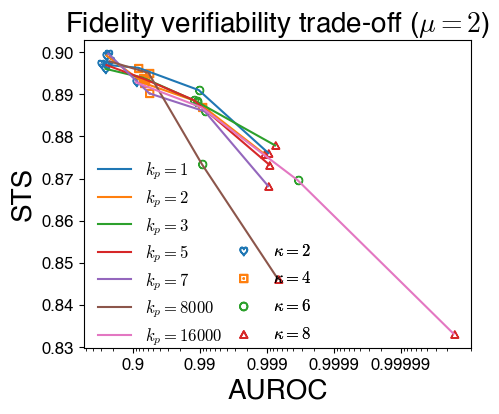}
}
\vskip -0.1in
\caption{Fidelity and verifiability for different $\id$, $\orthkid$ and $\str$}
\label{fig:STS_AUROC_diff_seed}
\end{figure}

We observe that different values for $\id$ does not result in noticeable impact on the fidelity and verifiability of the watermarked text, as shown in \cref{fig:STS_AUROC_diff_seed}.
Varying $\orthkid$ results in minor variations in fidelity and verifiability at high $\str$, but the pareto-front of the fidelity verifiability trade-off is similar across the different $\orthkid$. Clients using different $\orthkid$ could adjust the value of $\str$ to suite their requirements for fidelity and verifiability.

\subsection{Scalability}\label{sec:scalability_appendix}

We examine the scalability of \algname and benchmarks \algmbit, \algpnlw in practice by watermarking with different IDs and verifying with different IDs.

\subsubsection{Scalability when verifying with different IDs}\label{sec:scalability_appendix_verify}
\begin{figure}[ht]
\resizebox{\columnwidth}{!}{
\centering\includegraphics[height=3cm]
{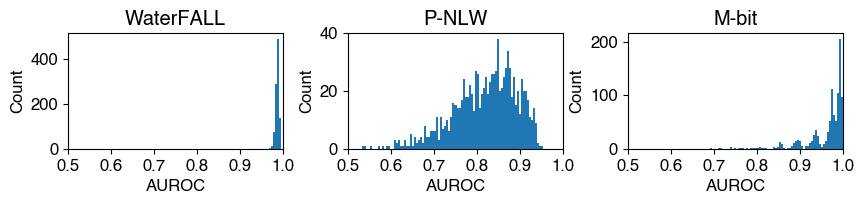}
}
\vskip -0.1in
\caption{
AUROC of $\swi{i}$ when verifying with $\id_i$ vs. $\id_{j \neq i}$. 
\algname has consistently high verifiability for all 1000 $\id_{j \neq i}$, compared to benchmarks which have many $\id_{j \neq i}$ with poor verifiability.
}
\label{fig:article_scalability}
\end{figure}

Using a dataset of text watermarked with ID $\mu=i$, we compare the verifiability using the correct ID ($\vop(\mu_i, \swi{i})$) against verifiability using the wrong IDs ($\vop(\mu_{j\neq i}, \swi{i})$). \cref{fig:article_scalability} shows the histogram plot for the AUROC comparing the 2 verification scores ($\vop(\mu_i, \swi{i})$ versus $\vop(\mu_{j\neq i}, \swi{i})$) for the different methods.

Notice that the AUROC of \algname for the different IDs are all closely clustered around the high value of 0.985. However, the AUROC of benchmarks \algmbit and \algpnlw show a very large range, with some IDs showing very low AUROC down to 0.69 and 0.53 respectively.

\begin{figure}[ht]
\centering
\resizebox{0.66\columnwidth}{!}{
\includegraphics{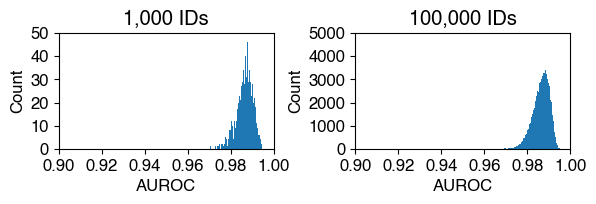}
}
\vskip -0.1in
\caption{Scalability of \algname for AUROC of $\swi{i}$ when verifying with $\id_i$ vs. $\id_{j \neq i}$, when using 1000 IDs versus 100,000 IDs.
Scaling up to 100,000 IDs shows the same narrow clustering of values around the high AUROC value of 0.985.
}
\label{fig:scalability_100k}
\end{figure}

To further support our claim of \algname having large scalability, we performed verification with 100,000 different IDs for \algname. \cref{fig:scalability_100k} shows that the distribution of AUROC values are similar when scaling up from 1,000 to 100,000 IDs, and this performance could be extrapolated into millions of IDs.

\subsubsection{Scalability when watermarking different IDs}\label{sec:scalability_appendix_watermark}
\begin{figure}[ht]
\centering
\resizebox{0.33\columnwidth}{!}{
\includegraphics{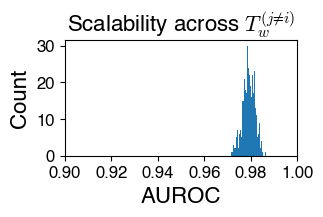}
}
\vskip -0.1in
\caption{AUROC of $\swi{i}$ vs. $\swi{j \neq i}$ when verifying with $\id_i$. \algname shows consistently high AUROC when verifying $\swi{i}$ with $\id_i$ compared to verifying $\swi{j \neq i}$ with $\id_{i}$}
\label{fig:scalability_many_t_w}
\end{figure}

We further explore the scalability of \algname when verifying text watermarked with different IDs. 
We compare the verifiability using the correct ID ($\vop(\mu_i, \swi{i})$) against verifiability using the wrong IDs ($\vop(\mu_i, \swi{j\neq i})$).
Due to the higher computational cost of watermarking compared to verification, we performed this experiments over a smaller subset of 358 pieces of text of the \texttt{c4 realnewslike} dataset. 500 different IDs were used to watermark the dataset. \cref{fig:scalability_many_t_w} shows the distribution of AUROC comparing the 2 verification scores $\vop(\mu_i, \swi{i})$ versus $\vop(\mu_{i}, \swi{j\neq i})$ for \algname is closely clustered around 0.98, similar to the results in \cref{sec:scalability_appendix_verify}. Note that a smaller number of text are considered for this experiment, resulting in the slightly difference in distribution.

\subsubsection{Discussion on scalability in practice}\label{app:benchmark_scalability}
\algmbit, \algpnlw suffer from poor scalability in practice, as shown above. As we consider watermarking or verification with the wrong ID $\mu_{j \neq i}$, there can be situations where the wrong ID differ from the correct ID at only 1 single bit, or very few bits. If the text is too short to be able to encode sufficient number of bits to include the differing bits, the watermarking method would be unable to differentiate between the 2 IDs during verification.

Even if the texts are sufficiently long, IDs that have few differing bits will be harder to differentiate. As discussed in \cref{app:benchmark_weakness}, errors could be present in the verification of watermark with \algmbit and \algpnlw. Such errors could overshadow the small differences in the watermarking and verification IDs, resulting in poor verification performance. To achieve satisfactory performance, \algmbit and \algpnlw would have to limit their scheme to IDs with sufficient number of differing bits, which further limit the scalability of their schemes.

On the other hand, \algname is not susceptible to such issues. As the watermark signal is not embedded directly into the specific substitutions in the text space, but rather into signals in the permuted token space determined by a hash of the ID, small differences in the ID results in drastically different permutations in the token space, and they are extremely unlikely to collide, i.e., 2 different IDs are extremely unlikely to map to the same permutations over the entire piece of text. As a result, \algname can achieve significantly higher scalability than \algmbit and \algpnlw in practice.

\subsection{Extraction}\label{app:extraction}
To evaluate extraction accuracy, we applied \cref{alg:extract} on the watermarked text. The accuracy is calculated based on the percentage of exact matches (extracted $\hat{\orthkid}$ matches the $\orthkid$ used to watermark the text).

Note that as there are 31999 supported $\orthkid$ when using the Fourier basis functions with \llama as the paraphraser, the probability of randomly guessing the correct $\orthkid$ is $\frac{1}{31999} = 0.003125\%$. Despite this, \algname is able to achieve high extraction accuracy of 48\% when extracting from a single text for our default setting of $\str=6$. This performance can be further improved when more pieces of watermarked text are available, such that accuracy improves to the high value of 99\% with only 5 pieces of text. This is done by combining multiple pieces of text watermarked by the same ID $\id$ and perturbation key $\orthkid$, by simply summing the cumulative token counts in $V_w$ space, $C$, of the different pieces of text, before performing step 7 of \cref{alg:extract}.

\section{Experimental details and additional results for attacks} \label{app:attacks}

\subsection{$\attackspace_1$} \label{app:a1}
\label{sec:word_attack}
Following \citet{kamaruddinReviewTextWatermarking2018}, we design three types of attack: insertion, deletion, and synonym substitution attacks for $\attackspace_1$. Attack strength indicates the rate of attacked words over the total number of words in a given content.

\textbf{Insertion attack.} We consider two types of insertion attacks mentioned in \citet{kamaruddinReviewTextWatermarking2018}:

(1) Localized insertion: this kind of attack inserts a random word into the original content at a random position. This is labeled as ``local'' in \cref{fig:insertion}.

(2) Dispersed insertion: multiple random words are added in multiple random positions into the original content. In our experiment, we iteratively insert a random English word into a random position of the original content.

\textbf{Deletion attack.} Random words are deleted, to attempt to distort the watermark in the original content.

\textbf{Synonym substitution attack.} Given original content, the synonym substitution attack tries to replace some words with their synonyms. In our experiments, we use the Natural Language Toolkit (NLTK) \citep{BirdKleinLoper09_nltk} to find a set of synonyms for a certain word, then choose a random word in this synonym set to replace the original word. We used the random function in the NumPy library \citep{harris2020array_numpy} to randomly select words to be substituted for these types of attacks.

\subsection{$\attackspace_2$} \label{app:a2}
\textbf{Translation attack} was performed with \texttt{gpt-3.5-turbo-0613}, with the following prompts, where the {language} field is ``Spanish'' and ``English''.

\noindent
\begin{minipage}{\textwidth}\begin{lstlisting}
{
    'role': 'system',
    'content': 'Translate the provided piece of text to {language}.'
}
{
    'role': 'user', 
    'content': '{text}'
}
\end{lstlisting}\end{minipage}

\textbf{Paraphrase attack} was performed with \llama, prompted in the following format.

\noindent
\begin{minipage}{\textwidth}\begin{lstlisting}
[INST] <<SYS>>
Paraphrase the user provided text while preserving semantic similarity. Do not include any other sentences in the response, such as explanations of the paraphrasing. Do not summarise.
<</SYS>>

{text} [/INST] 

Here is a paraphrased version of the text while preserving the semantic similarity:
\end{lstlisting}\end{minipage}

We ran further experiments using different LLMs to perform paraphrasing attack. The robust verifiability of \algname, \algmbit and \algpnlw are reported in \cref{tab:attack_2_diff_LLMs}. \algname achieves significantly higher robust verifiability than the benchmarks under paraphrasing attack across the different LLMs.

\begin{table}[ht]
\caption{Robust verifiability under paraphrasing attack with different LLMs.}
\centering
\begin{small}
\setlength\tabcolsep{2.2pt}
\begin{tabular}{r|cccc}
\toprule
  & \texttt{gemma-7b-it}\tablefootnote{https://huggingface.co/google/gemma-7b-it} & \texttt{Llama-2-7b-chat-hf}\tablefootnote{https://huggingface.co/meta-llama/Llama-2-7b-chat-hf} & \texttt{Mixtral-8x7B-Instruct-v0.1}\tablefootnote{https://huggingface.co/mistralai/Mixtral-8x7B-Instruct-v0.1} & \texttt{gpt-3.5-turbo} \\
  \midrule
\algname & \textbf{0.880}& \textbf{0.881}& \textbf{0.701} & \textbf{0.760} \\
\algmbit & 0.524       & 0.509              & 0.522 & 0.385       \\
\algpnlw & 0.374       & 0.359              & 0.467 & 0.512      \\
\bottomrule
\end{tabular}
\label{tab:attack_2_diff_LLMs}
\end{small}
\end{table}

\subsection{$\attackspace_3$} \label{app:a3}

We show the results of $\attackspace_3$ overlap watermark on \algname when the watermark overlap was applied on $\id$ or $\orthkid$ in \cref{tab:additional_overlap}. We can see that \algname can achieve high robust verifiability for overlap attack for both applications.
\begin{table}
    \centering
    \caption{Robust verifiability under overlap watermarking attack with different $\id$ or $\orthkid$.
    }
\begin{small}
    \begin{tabular}{l|cc}
\toprule
        & Pre-attack & Post-attack\\
         \midrule
        Overlap $\id$ & 0.992 & 0.815\\
        Overlap $\orthkid$ & 0.992 & 0.743\\
\bottomrule
    \end{tabular}
\end{small}
\vskip -0.15in
    \label{tab:additional_overlap}
\end{table}

\begin{figure}[ht]
\centering
\resizebox{0.75\columnwidth}{!}{
\centering\includegraphics{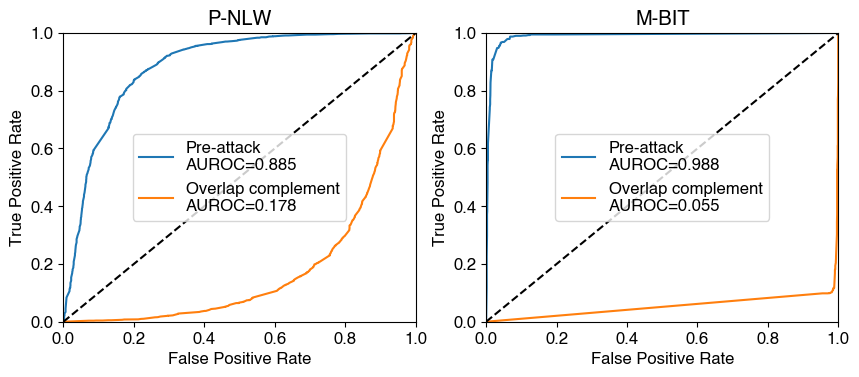}
}
\vskip -0.1in
\caption{
ROC curves and corresponding AUROC values of $\attackspace_3$ with the complement of binary watermark key of \algpnlw and \algmbit.
}
\label{fig:overlap_com}
\end{figure}

\paragraph{$\attackspace_3$ on benchmarks with complement binary key} \label{sec:complement_attack}
We consider the worst-case scenario of robust verifiability under $\attackspace_3$ for two traditional approaches \algpnlw and \algmbit. Because these two methods are based on embedding binary keys in the watermarking stage, we try to apply $\attackspace_3$ with the complement of the binary watermark key that was extracted as part of the verification process (replacing bit 0 with bit 1 and vice versa), to illustrate the worst-case scenario. We conduct this experiment with setting as \cref{sec:exp_article}. The results are illustrated in \cref{tab:overlap_com} and \cref{fig:overlap_com}. Do note that attacks could engineer their attacks by performing overlap watermarking with a mixture of watermark bits, random bits and complement bits, to target any AUROC value between the pre-attack and overlap complement AUROC.

\begin{table}
    \caption{AUROC of \algpnlw and \algmbit under $\attackspace_3$ with the complement of binary watermark key (worst case scenario)}
    \centering
    \begin{small}
    \begin{tabular}{c|cc}
    \toprule
         &  Pre-attack& Overlap complement\\
         \midrule
         \algpnlw &  0.8848& 0.1780\\
         \algmbit &  0.9882& 0.0547\\
         \bottomrule
    \end{tabular}
    \end{small}
    \label{tab:overlap_com}
\end{table}

\subsection{$\attackspace_4$} \label{app:a4}
To perform the in-context prompting experiments, we made use of \texttt{gpt-3.5-turbo-1106} to generate 3 questions each for 300 text articles. The following prompt was used to generate the questions.

\noindent
\begin{minipage}{\textwidth}\begin{lstlisting}
{
    'role': 'system',
    'content': 'Using the provided article, create 3 reading comprehension questions.'
}
{
    'role': 'user', 
    'content': '{text}'
}
\end{lstlisting}\end{minipage}
We then separately prompt \texttt{gpt-3.5-turbo-1106}, providing the watermarked text as the context to answer the questions. 

\noindent
\begin{minipage}{\textwidth}\begin{lstlisting}
{
    'role': 'system',
    'content': 'Using the provided article, answer the questions.'
}
{
    'role': 'user', 
    'content': '{text}\n\n{questions}'
}
\end{lstlisting}\end{minipage}

\subsection{Additional results for robust verifiability}
Beyond AUROC reported in the main paper, we additionally report the true positive rate (TPR) at fixed false positive rate (FPR) of 0.1 and 0.01 for verifiability and robust verifiability under different attacks across different watermarking methods in \cref{tab:tpr_fpr}.

\begin{table}[ht]
\caption{TPR at FPR of 0.1 and 0.01 for verifiability and robust verifiability.}
\centering
\begin{small}
\begin{tabular}{l|l|l|llll}
\toprule
FPR                  &          & Pre-attack     & $\attackspace_{2-T}$ & $\attackspace_{2-T}$ & $\attackspace_{3}$ & $\attackspace_{4}$ \\
\midrule
\multirow{3}{*}{0.1} & \algname & 0.982          & \textbf{0.890}       & \textbf{0.750}       & \textbf{0.640}     & \textbf{0.472}     \\
                     & \algpnlw & 0.667          & 0.078                & 0.110                & 0.281              & 0.114              \\
                     & \algmbit & \textbf{0.993} & 0.126                & 0.126                & 0.520              & 0.000              \\
\midrule
\multirow{3}{*}{0.01} & \algname & \textbf{0.910} & \textbf{0.608} & \textbf{0.405} & \textbf{0.284} & \textbf{0.122} \\
                     & \algpnlw & 0.110          & 0.007                & 0.010                & 0.037              & 0.032              \\
                     & \algmbit & 0.693          & 0.126                & 0.000                & 0.126              & 0.000              \\
\bottomrule
\end{tabular}
\label{tab:tpr_fpr}
\end{small}
\end{table}

Note that under \algname, we are able to drastically improve the verification performance when multiple pieces of text are available to be considered, where a realistic setting would involve multiple samples from the adversaries that we could test the watermarks for. In reality, IP holders are concerned about large-scale unauthorized IP use (i.e., multiple infringements) rather than one-off cases.

To demonstrate this, we ran an experiment where we test our watermarks given multiple samples under attack $\attackspace_4$. Desipte the low TPR of 0.472 and 0.122 for FPR of 0.1 and 0.01 respectively when only considering 1 sample, our results demonstrates that given just 10 samples, we are able to achieve a TPR of 0.907 even with the strict requirement of a FPR of 0.01.
The TPR increases to even 1.000 given 17 samples when we have the requirement of 0.1 FPR. This is also realistic because in practice, IP holders may use this as a screening tool for suspicious parties, to investigate them further, and hence would be alright with a higher FPR.

\section{\algname in code watermarking}\label{app:code_app}
\subsection{Code watermarking experiment settings}
In the main paper, we report the result of code watermarking on the MBJSP dataset \citep{mbxp_athiwaratkun2022} with the data ownership problem setting. This is a JavaScript dataset including around 800 crowd-sourced JavaScript programming problems. To show the ability of \algname on watermarking other programming languages, we also perform data ownership watermarking on Python datasets, which can be found in \cref{app:code_python}. 

In this setting, we use 
\texttt{Phind-CodeLlama-34B-v2}\footnote{https://huggingface.co/Phind/Phind-CodeLlama-34B-v2}
, as LLM paraphraser for code watermarking, the square wave basis with $\orthkid=1$ (\cref{sec:orthgonal_fn_ex}) for watermark perturbation and randomly choose $\id=10$ in all code experiments.
As default, we denote \algname~code to indicate \algname in this code watermarking settings. 
Moreover, we also show that prompt engineering techniques, such as Reflexion \citep{reflexion} could improve the fidelity of watermarked code while preserving the verifiability (\cref{app:code_rf}).
For \algsrcmk \cite{srcmarker}, we configured their algorithm for 16-bit watermarks, to demonstrate scalability of at least $10^5$.

\begin{figure}[ht]
    \centering
    \resizebox{0.4\columnwidth}{!}{
    \includegraphics{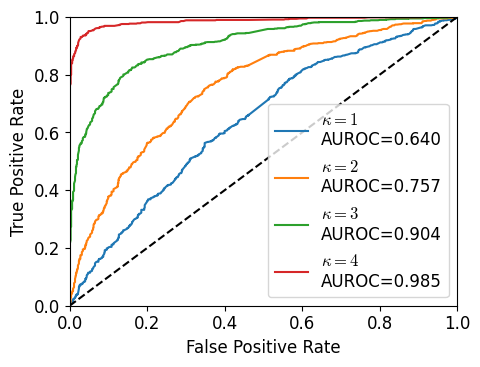}
    }
    \vskip -0.1in
    \caption{The ROC curves and corresponding AUROC values on the MBJSP dataset using \algname~code.}
    \label{fig:code_auc_curve}
\end{figure}

For verifiability evaluation, we use the same evaluation protocol as article watermarking in \cref{sec:exp_article}.
As a result, the ROC curve and AUROC values for \algname~code are shown in \cref{fig:code_auc_curve} 

\paragraph{Watermarked code fidelity evaluation}
As mentioned in the main paper, we evaluate the fidelity of the watermarked code by evaluating its accuracy based on functional tests for the original code and use the standard pass@k metric \citep{passk_metric, chen2021codex} for evaluating functional correctness. Given the deterministic nature of the baseline \algsrcmk \citep{srcmarker}, which inherently upholds fidelity, the pass@10 metric is adopted to facilitate a fair comparison between \algname and \algsrcmk in terms of fidelity performance. This metric specifically measures the likelihood of \algname producing watermarked code that passes unit tests within 10 generation attempts.  The pass@10 metric is also realistic in practice as it aligns with real-world scenarios where clients can assess the quality of watermarked code through predefined tests and subsequently regenerate the code if test failures arise.

To evaluate the functional correctness of code, we adapt the JavaScript evaluation protocol from \citet{mbxp_athiwaratkun2022} for the MBJSP dataset. On the other hand, for Python evaluation, we adapt the HumanEval \citep{chen2021codex} code evaluation protocol\footnote{https://github.com/openai/human-eval} and test script from both datasets \citep{chen2021codex, MBPP}. However, the watermarked code usually modifies the original function name into some related names, so we use Levenshtein distance to find the new text function in the watermarked code. For a more precise evaluation of the watermarked code, this related function name-finding process can be improved by using other similarity distances, such as the Semantic Textual Similarity (STS) score.

\subsection{\algname code + Reflexion methodology} \label{app:code_rf}
In this section, we show that some prompt engineering approaches could help the watermarked code improve fidelity without hurting the verifiability. Adapting the techniques from \citet{reflexion}, we try to correct the watermarked code through the LLM-based self-reflection mechanism. After being watermarked with \algname code, this watermarked code undergoes a correcting process via multiple feedback loops (3 feedback loops in our experiments). Each feedback loop contains two self-reflection components aiming to perform syntax correction and functional similarity alignment. Each self-reflection component performs two main steps: 1) evaluating or analyzing the given information based on task criteria, e.g., the correctness of programming syntax. 2) regenerate the “better” code based on given feedback. 

Applying the same LLM in \algname code to the self-reflection component plays a crucial role in this combination. This is simply because LLM is a good way to handle and generate linguistic feedback, which contains more information than scalar results in the evaluation step. Moreover, watermarking LLM helps the final code preserve the robust and scalable watermark signal through the correction step, which is the ultimate goal of our text watermarking framework. The prompts to perform the syntax correction step and functional similarity alignment are illustrated in \cref{app:code-prompt-app}. 

\begin{figure}[ht]
\centering
\resizebox{0.4\columnwidth}{!}{
\includegraphics{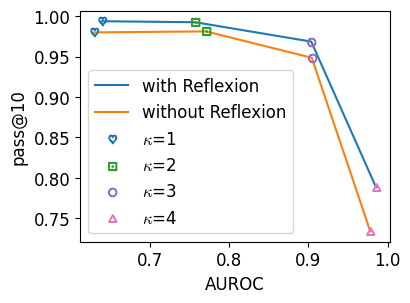}
}
\vskip -0.1in
\caption{The effect of Reflexion in \algname code on MBJSP dataset}
\label{fig:code_rf_bt}
\end{figure}

The effect of the Reflexion approach is shown in \cref{fig:code_rf_bt}. From this illustration, we can see that Reflexion improves fidelity while maintaining high verifiability of \algname code. So we apply this technique in all code watermarking experiments. 

\subsection{Verifiability and fidelity trade-off}
\begin{figure}[ht]
\centering
\resizebox{0.4\columnwidth}{!}{
    \centering\includegraphics{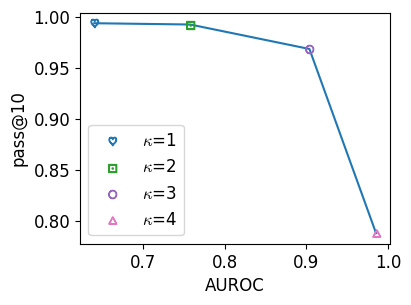}}
\caption{Verifiability and fidelity trade-off of \algname~code on the MBJSP dataset}
\label{fig:code_tradeoff}
\end{figure}

\cref{fig:code_tradeoff} shows the trade-off of verifiability and fidelity can be adjusted via $\str$. Similar to article watermarking in \cref{sec:exp_article}, increasing watermark strength $\str$ can increase verifiability but lower fidelity. Therefore, the users can adjust $\str$ to balance the trade-off based on their preference.

\subsection{Scalability of \algname in code watermarking}
One of the advantages of \algname over baseline \algsrcmk is in terms of scalability. \algsrcmk verifiability depends heavily on the number of watermarked bits (scalability), larger number of bits, worse verifiability \citep{srcmarker}. Therefore, to ensure high verifiability, \algsrcmk can not support larger scalability. In contrast, the verifiability of \algname is independent to its scalability, and this scalability only depends on the vocabulary size of the tokenizer. In our experiments (\cref{tab:code_table}), we use 
\texttt{Phind-CodeLlama-34B-v2}
, which has a large vocabulary size as same as \llama, which $M \sim 10^{130274}$, far better than $M \sim 10^5$ of \algsrcmk 16-bits.

\subsection{\algname in watermarking Python code} \label{app:code_python}
Inheriting the multi-lingual ability of LLM, \algname can easily apply to new programming languages without the need for pre-defined syntax rules. This is a big advantage of \algname in comparison to AST-based code watermarking approaches like \algsrcmk \citep{srcmarker}. We show that \algname can also watermark Python code, through experiments on the MBPP dataset \citep{MBPP} which includes around 1000 crowd-sourced Python programming problems.
We show the verifiability and fidelity results of \algname on watermarking Python code in \cref{tab:code_python}.

\begin{table}[ht]
    \centering
    \begin{small}
        \begin{tabular}{c|c|c}
            \toprule
                 &pass@10&AUROC  \\
            \midrule
                 MBJSP&0.969&0.904 \\
                 MBPP&0.954&0.897 \\
            \bottomrule
        \end{tabular}
    \end{small}
    \caption{\algname~code achieves high verifiability and fidelity on MBJSP and MBPP datasets.}
    \label{tab:code_python}
\end{table}

\subsection{LLM prompts for code watermarking}\label{app:code-prompt-app}
We use the following prompts and apply the chat template of 
\texttt{Phind-CodeLlama-34B-v2}, which follows the alpaca instruction prompt format
 on these prompts.

\noindent
\begin{minipage}{\textwidth}
\textbf{Code paraphrasing}
\begin{lstlisting}
### System Prompt
You are given a user-provided code snippet.
Please do ONLY two tasks:
1. Refactor the provided code snippet with the following requirements:
- retain all imported libraries.
- keep the same programming language.
- retain the function names and functionality of the code.
- don't complete the code, just refactor it.
- don't explain.
2. Return the response with the refactored code snippet in the following format strictly:
```
<refactored code>
```
Do not generate any comments or explaining texts.
### User Message
```
{input code}
```
### Assistant
Here is the refactored code:
```
\end{lstlisting}\end{minipage}

\noindent
\begin{minipage}{\textwidth}
\textbf{Functional similarity alignment}
\begin{lstlisting}
### System Prompt
You are given two code snippets, code A and code B. Modify code B based on code A, such that these two code have the same functionality, input, and output. Return the response with corrected code B in the following format strictly:
```
<corrected code B>
```
Do not generate any comments or explaining texts.
### User Message
code A:
```
{original code}
```

code B:
```
{watermarked code}
```
### Assistant
Here is the code B:
```
\end{lstlisting}\end{minipage}

\noindent
\begin{minipage}{\textwidth}
\textbf{Code syntax correction}
\begin{lstlisting}
### System Prompt
Double-check the code to make sure the syntax is correct. Only generate the corrected code in the following format.
```
<corrected code>
```
Do not generate any comments or explaining texts.
### User Message
```
{watermarked code}
```
### Assistant
Here is the corrected code:
```
\end{lstlisting}\end{minipage}

\subsection{Watermarked code examples}
\begin{figure}[ht]
    \centering
    \resizebox{0.95\columnwidth}{!}{
    \includegraphics{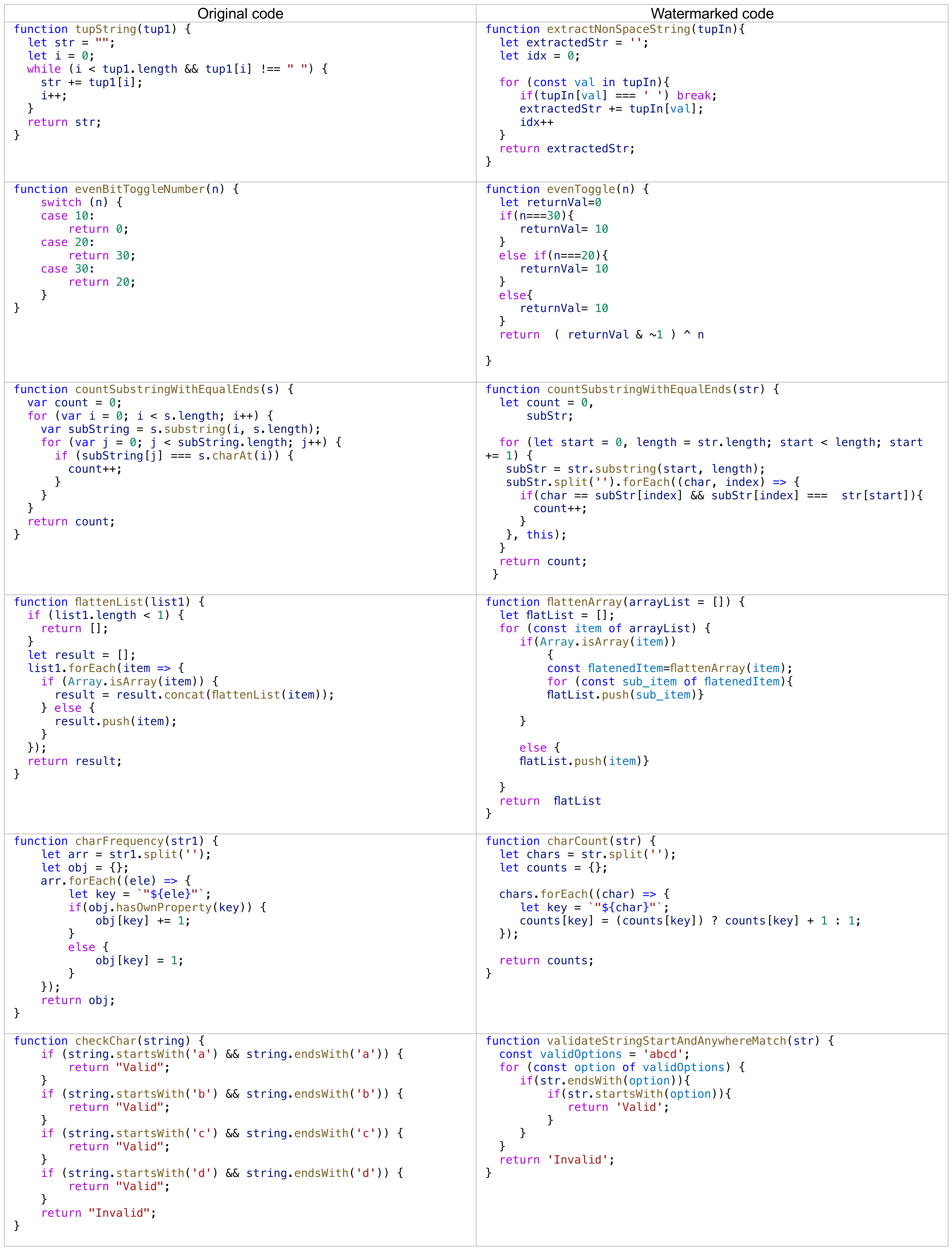}
    }
    \caption{Example of watermarked code with \algname. \algname~code changes not only the variable names but also the ways of representing the same code logic (e.g., ternary operator vs. conditional statement), which results in high verifiability while preserving code functionality (high fidelity).}
    \label{fig:code_example}
\end{figure}

Examples of code watermarking by \algname are illustrated in \cref{fig:code_example}. Note that \algname~code changes not only the variable names but also the ways of representing the same code logic, which results in high verifiability while preserving high fidelity. 

\section{Details of experiments on LLM data provenance}\label{app:provenance}

\subsection{LLM fine-tuning experimental setup}\label{sec:lora_setup}
To fine-tune the 1.5B parameter \texttt{gpt2-xl} model, we used the LoRA framework \citep{hu2022lora}, with LoRA rank of 16 and target modules \texttt{c\_attn, c\_proj, c\_fc}. The models were fine-tuned for a total of 5 epochs, with default batch size of 128 and learning rate of 0.0003. Fine-tuning was performed on a single Nvidia L40 GPU, requiring an average of approximately 15 minutes per client (4000 data samples for each client) for the fine-tuning of the model.

\subsection{Verifiability of watermark in the model fine-tuned over watermarked text}\label{sec:attribution_verifiability}
To evaluate the verifiability of the watermark, we prompted the fine-tuned model with randomly selected abstracts from the training set used to fine-tune the model. We truncated the abstracts to the first 50 tokens, which is supplied to the model without any other additional prompts, for the model to generate completions to the input. We limited the generation to a maximum of 100 newly generated tokens. Note that in real applications, generating more tokens could improve the verifiability performance, as we have demonstrated in the data ownership experiments.
Only the generated tokens were considered when evaluating the verifiability of the watermark. The same ID and $\orthkid$ used during watermarking was used to perform verification. For the model fine-tuned on the original unwatermarked text, the corresponding ID and $\orthkid$ that was used for the watermarked text was used for verification.

\subsection{Fidelity of model fine-tuned over watermarked text}\label{sec:attribution_fidelity}
We used \texttt{lm-evaluation-harness}\footnote{https://github.com/EleutherAI/lm-evaluation-harness} \citep{eval-harness} to evaluate the fine-tuned models for its fidelity over several different datasets \citep{gao2020pile,merity2016pointer,wang2018glue,dolan2005automatically,bisk2020piqa,levesque2011winograd}. \cref{tab:alpaca_fidelity} reports the models fine-tuned over the watermarked datasets results in minimal differences in fidelity as compared to the model fine-tuned over the unwatermarked datasets. This shows that act of watermarking data used for fine-tuning does not significantly affect its value for fine-tuning.

\begin{table*}[ht]
    \caption{Fidelity of model fine-tuned using watermarked text (Watermarked) and unwatermarked text (Unwatermarked) of different number of clients $M$, evaluated over the various datasets.}
    \centering
    \begin{small}
    \begin{tabular}{rr|cccccc}
    \toprule
         \multirow{2}{*}{Dataset}& & \multicolumn{5}{c}{$M$}\\
                                 & & 1 & 5 & 10 & 20 & 100\\
         \midrule
         \multirow{2}{*}{Pile-ArXiv (ppl)} & Watermarked   & 2.209 & 2.218 & 2.218 & 2.180 & 2.166\\
                                           & Unwatermarked & 2.192 & 2.210 & 2.197 & 2.170 & 2.154\\
         \midrule
         \multirow{2}{*}{Wikitext (ppl)}   & Watermarked   & 1.771 & 1.770 & 1.780 & 1.787 & 1.818\\
                                           & Unwatermarked & 1.766 & 1.769 & 1.774 & 1.783 & 1.814\\
         \midrule
         \multirow{2}{*}{MRPC (acc)} & Watermarked   & 0.662 & 0.618 & 0.674 & 0.581 & 0.326\\
                                     & Unwatermarked & 0.679 & 0.627 & 0.627 & 0.380 & 0.314\\
         \midrule
         \multirow{2}{*}{PIQA (acc)} & Watermarked   & 0.687 & 0.676 & 0.682 & 0.676 & 0.673\\
                                     & Unwatermarked & 0.686 & 0.682 & 0.683 & 0.680 & 0.678\\
         \midrule
         \multirow{2}{*}{WNLI (acc)} & Watermarked   & 0.563 & 0.620 & 0.535 & 0.549 & 0.493\\
                                     & Unwatermarked & 0.620 & 0.577 & 0.592 & 0.563 & 0.535\\
         \bottomrule
    \end{tabular}
    \end{small}
    \label{tab:alpaca_fidelity}
\end{table*}

\section{Adapting model watermarking schemes into \algname framework}
\label{sec:wtf_kgw}
There exists a separate area of research addressing a different problem setting of model watermarking, where instead of watermarking existing text, newly generated text from LLMs are watermarked. Contrary to the setting of text watermarking, where scalability is a critical requirement, model watermarking schemes are only concerned with a single client (the LLM provider).

Despite this, we could try adapting some model watermarking schemes into the \algname framework, though some features of our framework may not be achievable. One such possible scheme that can be adapted is KGW \citep{kirchenbauerWatermarkLargeLanguage2023}. To adapt, KGW, line 5 and 6 of \cref{alg:watermark} would be replaced with "Green" and "Red" lists, with $\gamma=0.5$. In order to satisfy the scalability criteria, we appended our watermark ID $\id$ to the hash of the previous token, to be used to seed the random partition of the vocabulary list into "Green" and "Red" lists. For verification, we used $z$-score as proposed in their paper.

Despite our various additions to the scheme (such as increasing its scalability by adjusting the original function for seeding the random partitioning), this \algname variant under performs compared to our original proposed \algname implementation, and is still missing key features such as the ability for clients to embed and extract metadata from text after verification with their ID.

\begin{figure}[H]
\centering\resizebox{0.4\columnwidth}{!}{
\includegraphics{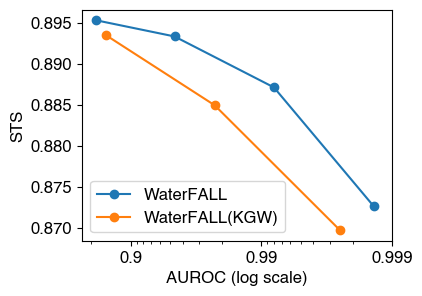}
}
\vskip -0.1in
\caption{
Strictly better fidelity-verifiability Pareto frontier for \algname(Ours) than \algname(KGW).
}
\label{fig:pareto_KGW}
\end{figure}

\cref{fig:pareto_KGW} shows that \algname(Ours) has a strictly better fidelity-verifiability Pareto frontier, i.e., for any required fidelity (STS score), \algname(Ours) has higher verifiability than \algname(KGW).

We also performed comparison of robust verifiability for \algname(Ours) vs. \algname(KGW). For fair comparison, the watermark strength was selected such that the STS score were similar for both variants (\algname(Ours): 0.887; \algname(KGW): 0.885). \cref{tab:robust_KGW} shows that due to better Pareto frontier of \algname(Ours), we are able to achieve a higher verifiability both before and after attacks, with the watermarked texts at the same fidelity as \algname(KGW).

\begin{table} [ht]
\vskip -0.1in
    \centering
    \caption{
    \algname(Ours) has better robust verifiability than \algname(KGW).
    }
\begin{small}
    \begin{tabular}{lc|cccc}
\toprule
         &  Pre-attack&  $\attackspace_{2-T}$&  $\attackspace_{2-P}$& $\attackspace_3$\\
\midrule
         \algname(Ours)& \textbf{0.992} & \textbf{0.951} & \textbf{0.881} & \textbf{0.815}\\
         \algname(KGW)& 0.977& 0.915& 0.811& 0.718\\
\bottomrule
    \end{tabular}
\end{small}
    \label{tab:robust_KGW}
\end{table}

\section{Differences with model-centric watermarking}\label{app:comparison_model_centric}

Our paper focuses on text watermarking, where our problem setting (\cref{sec:setting}) is on watermarking existing text (e.g., containing IP) produced by many clients (with any method including human written), such that each client can verify text that were watermarked with their own unique watermark, and additionally ensure that the watermark is robust to attacks and downstream uses by other LLMs (e.g., prompting, fine-tuning).

On the other hand, there exists a separate line of work focusing on a different problem of model-centric watermarking, which marks output from these watermarked models (e.g., differentiate text generated by these LLMs vs. that by humans).

The problem settings of such model-centric watermarking considers a specific LLM, and addresses how to design an algorithm that allows distinguishing the output of that specific LLM from other text (e.g., human generated). 
In this setting, the scalability issue is ignored, as only 1 client (the LLM provider) is considered. Additionally, LLM watermarking does not watermark individual original texts, and hence do not have the challenging requirements of preserving semantic content of these original texts. Rather, it typically only considers generative text quality through metrics like perplexity. Therefore, LLM watermarking methods tackles a different problem and should not be confused with the focus of our work.

To provide more detailed comparison on the differences with our work, we further separate model-centric watermarking into the following classifications:

\begin{enumerate}
    \item \emph{Text watermarking of text} generated from black-box LLMs.
    \item \emph{White-box LLM watermarking} leading to generated text which contains the model's watermarks.
    \item \emph{Black-box LLM watermarking} such that a watermarked model's output is passed to black-box models, with outputs that are still watermarked.
\end{enumerate}

\subsection{Text watermarking of text generated from black-box LLM}

To the best of our knowledge, the only works related to this topic we have found so far are the unpublished works \citep{yang2023watermarkingblackbox} and \citep{chang2024postmark}.

\citet{yang2023watermarkingblackbox} applies text watermarking methods to the specific use case of text generated by black-box language models and is therefore essentially a text watermarking paper. The text watermarking method of \citet{yang2023watermarkingblackbox} is similar to the M-BIT benchmark \citep{yoo-etal-2023-robust} that we considered in the main paper, and essentially encodes watermarks by first identifying words to replace (based on linguistic rules), then finds synonyms for them which are used to represent bits of the watermarking signal. Although the two methods differ in the way of selecting which word to perform watermarking (sentence/word embedding similarity for \citet{yang2023watermarkingblackbox} and a 2nd BERT model for M-BIT), given their similar characteristics, both methods ultimately still suffer from robust verifiability compared to \algname.

Nonetheless, we have performed additional experiments with their method on the same \texttt{c4-realnewslike} dataset from our paper, and considered the attacks $A_2$ and $A_3$. Note that \algname has significantly higher robust verifiability compared to \citet{yang2023watermarkingblackbox}, similar to its better performance over the other benchmarks M-BIT and P-NLW.

\begin{table}
    \centering
    \caption{
    Comparison of robust verifiability of \algname versus \citet{yang2023watermarkingblackbox}
    }
    \vskip 0.15in
\begin{small}
\setlength\tabcolsep{3.5pt}
    \begin{tabular}{lc|ccc}
\toprule
         & Pre-attack & $\attackspace_{2-T}$ & $\attackspace_{2-P}$& $\attackspace_3$\\
\midrule
         \algname& \textbf{0.992} & \textbf{0.951} & \textbf{0.881} & \textbf{0.815}\\
         \citet{yang2023watermarkingblackbox}& 0.975 & 0.761& 0.659& 0.474\\
\bottomrule
    \end{tabular}
\end{small}
\vskip -0.15in
    \label{tab:advanced_attack_watermarkingblackbox}
\end{table}

\citet{chang2024postmark} is a concurrent unpublished work that applies the watermark to LLM generated text by prompting another LLM to insert a list of words into the text. The list of words are selected based on a pseudo-random mapping of text embedding to words and top-k words semantically similar to the text. Verification is performed by counting the proportion of words in the list that are present in the text. Robustness against paraphrasing is achieved by also including words from the text that are sufficiently semantically similar to the words in the list.

\subsection{White-box LLM watermarking}
This line of work assumes access to the model and directly changes the model generation process to embed the watermark, primarily to differentiate the text generated by specific LLMs vs. for example that by humans. This type of model watermarking that has become a rapidly growing field, especially since the proposal of the KGW watermark \citep{kirchenbauerWatermarkLargeLanguage2023}. Although these works eventually end up with (model-centric) watermarks in the output of LLMs which are also text, they are actually solving a different problem setting from our work. Our work is focused on watermarking any given text, rather than watermarking an LLM such that its output will all end up being watermarked.

Even though they are not directly comparable, as mentioned in the main paper, some of these white-box LLM watermarking works might be adapted as sub-routines of \algname if they meet our framework's requirements. We have run additional experiments to demonstrate this by introducing a new \algname framework implementation variant that swaps our watermarking scheme described in Sec. 3.3 with a modified KGW watermarking scheme, with changes to make it fit our framework, such as appending our watermark ID $\mu$ to the hash of the previous token, to be used to seed the random partition of the vocabulary list into "Green" and "Red" lists.

Despite our attempts to adapt the scheme (such as increasing its scalability by adjusting the original function for seeding the random partitioning), key features such as the ability for clients to embed and extract metadata from text after verification with their ID \cref{alg:extract} will not be available for this \algname variant.

We ran additional experiments to compare this \algname variant [\algname(KGW)] with our original watermarking scheme [\algname(Ours)] in \cref{sec:wtf_kgw}. \cref{fig:pareto_KGW} demonstrate that \algname(Ours) has a strictly better fidelity-verifiability Pareto frontier, i.e., for any required fidelity (STS score), \algname(Ours) has higher verifiability than \algname(KGW).

We also performed comparison of robust verifiability for \algname(Ours) vs. \algname(KGW). We can see that due to better Pareto frontier of \algname(Ours), with the watermarked texts at the same fidelity as \algname(KGW), we are able to achieve a higher verifiability both before and after attacks.

Another type of white-box LLM watermarking introduced in \citet{wang2023wasa} modifies the model structure, and fine-tune the LLM to generate invisible Unicode watermark tokens during the generation process. This setting is extremely restrictive, as only the particular fine-tuned watermark model trained to generate the watermark tokens can be used. Additionally, similar to other Unicode watermarking methods, the watermark can be easily erased, rendering the method ineffective.

\subsection{Black-box LLM watermarking}
This line of work considers how to ensure that text generated from a client-controlled LLM may be watermarked such that other black-box models (e.g., neural networks) owned by adversaries that rely on the watermarked LLM would also have their output watermarked. Similar to "white-box LLM watermarking" described above, the focus of these works are on watermarking the specific models in question, although the output of these models may be text, which are the channels in which the model watermarks are transferred. An example of these type of works would be \citet{Li_Cheng_Li_Du_Zhao_Liu_2023}, which clearly have methods specific to model-centric training and watermarking, and hence cannot be applied to text watermarking.

\section{Comparison with plagiarism checkers} \label{app:plagiarism_check}
Although tackling the similar issue of IP protection and plagiarism detection, works on plagiarism checkers tackle a distinctly different problem from our problem setting, and cannot be used in our problem setting.

Firstly, contrary to watermarking where a watermark signal is actively embedded into the text, traditional plagiarism detection depends on passive detection, typically via pairwise comparisons of a suspected text to a large corpus of reference text. In their setting, a single (or small number) of suspected text is to be examined for plagiarism. They accomplish this by maintaining a huge database of reference text, and each suspected text is compared pairwise to each piece of reference text.
Such pairwise comparison of the suspicious text with all possible reference text is extremely computationally expensive \citep{Plagiarism_Detection}. In our problem setting of identifying unauthorized usage of textual data, clients could desire to scan through the entire Internet’s worth of textual content for potential plagiarism, and the shear amount of data makes such techniques computationally infeasible. With watermarking, only the suspected text is required during the verification process, without requiring the reference text to be compared against.

Secondly, due to the requirement to maintain a huge database of reference text, which is costly for individual clients, this task is currently commonly subcontracted out to third party detection systems (e.g., Turnitin). These vendors can have unfavorably broad licensing agreements regarding texts that were submitted for checking \citep{deZwart2018}.
Such approaches are not feasible in situations where either the original reference data or the suspected text are sensitive and cannot be shared with these external vendors, greatly limiting the applications where plagiarism checker can be deployed in.

\section{Practical considerations for real world deployment of \algname} \label{app:practical_deployment}

\algname's initial setup and computational resources for large-scale applications are low and practically viable. This makes actual large-scale deployment of text watermarking feasible, which is currently not possible given the current state of the art (SOTA) watermarking methods' limitations and resource requirements.

We illustrate this by laying out two approaches (decentralized or centralized) to deploying \algname, both of which have low initial setup and computational cost requirements.

\subsection{Decentralized deployment}

In this approach, clients randomly generate their own IDs (given the large space of supportable IDs), and can do watermark and verification operations on their own using their laptops with minimal setup.

\paragraph{Setup}
For most common text types/languages supported by LLMs, clients could immediately run \algname with no setup, given a default LLM and \algname settings, to generate the watermarked text $\sw$.

\paragraph{Computational cost}
\algname's watermarking computational cost is just that of running inference of the LLM paraphraser, with negligible overheads. Using a GPU available in many laptops (Nvidia RTX 5000), a user could use the Llama-2-13b model to watermark a text in $<25$s to already achieve great performance, as shown in Table 2 in our paper. We expect that the cost of running high performance LLMs on personal devices (e.g., MacBooks, laptops with GPUs) will get cheaper and cheaper, given the rapidly evolving landscape of LLMs.

\algname's verification operation is extremely fast and can be run on just CPU ($<0.04$s per text), without the need for any LLM. For practical applications, the verification operation will be the main operation run multiple times, rather than the watermarking operation (typically only once before the user publishes the text). \algname's verification operator is 2-5 orders of magnitude faster than baseline text watermarking methods (Table 2 in our paper).

\subsection{Centralized deployment}
In this approach, central parties assigns clients unique IDs, and run the \algname watermarking and verification operations for them. This is similar to how LLM service providers are providing interfaces or APIs for LLM queries.

\paragraph{Setup}
At a minimum, they could do the same as individuals in the decentralized approach and not need to do any setup. However, given their scale, they could also provide customized service by optimizing the choice of LLMs and \algname settings for specific non-common text types or other user requirements (see \cref{app:adaptability} below for clarification on adaptability).

\paragraph{Computational cost}
Existing LLM service providers could easily provide this additional watermarking service to clients, given the minimal overheads of \algname over processing a single LLM chat API call.
The speed of our verification operation even allows companies to provide value-added services such as near-real-time scanning of newly-published articles from target sources to detect any plagiarism.

\subsection{Adaptability to different LLMs} \label{app:adaptability}

A key strength of \algname is that it evolves together with the evolving landscape of LLMs, with increasingly better watermarking performance as LLMs become more capable. As LLMs become more capable, they would be able to better preserve semantic meaning of the original text while still embedding watermarks when used as LLM paraphrasers via \algname. This allows \algname to achieve higher fidelity-verifiability Pareto frontier, and reduce any fidelity degradation while using higher watermarking strength for greater robust verifiability.

\begin{figure}[H]
\centering\resizebox{0.4\columnwidth}{!}{
\includegraphics{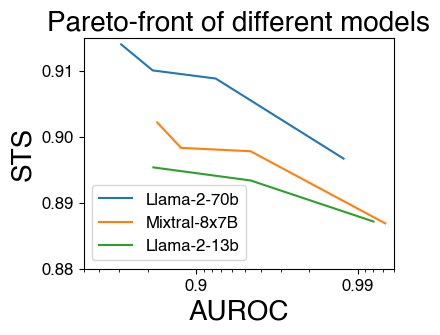}
}
\vskip -0.1in
\caption{
Plot of Pareto frontier of different LLMs, where larger/newer models show better Pareto fronts on the fidelity-verifiability trade-off.
}
\label{fig:pareto_diff_llm}
\end{figure}

To illustrate, we have performed additional experiments with other LLM models as paraphraser models, with the same \texttt{c4-realnewslike} dataset used in the main paper. \cref{fig:pareto_diff_llm} shows that the newer/larger models have higher Pareto fronts with higher STS scores for the same verifiability values. Going forward, we expect further significant improvements in LLM capabilities, allowing \algname's performance to also significantly improve.

\subsection{Selection of watermarking LLM and hyperparameter}
As with any adaptable methods, \algname would require some effort to gain boosted performance in specific domains (e.g., text type or language). That said, the \algname framework is designed to reduce such efforts, and it is relatively easy for a user to perform such fine-tuning given only 1 hyperparameter to tune (watermarking strength $\str$) and the choice of LLM paraphraser. For example, the user could just follow these simple steps:

\begin{enumerate}
    \item Identify the SOTA LLM for the domain, to use as the LLM paraphraser component. As a domain expert and content creator (of the text to be watermarked), the client should be familiar with what is available. Given the evolving landscape of LLMs, we believe that it is realistic for each domain to have a relatively capable fine-tuned model.
    \item Run \algname with default watermarking strength $\str$ and assess if the fidelity and robust verifiability of the text meets expectation. As a domain expert, the client can assess if the text has sufficient fidelity or use a domain-specific fidelity metric to automate the check. The client can also use an automated suite of robustness checks (comprising standard attacks) would assess the expected robust verifiability of the watermarked text.
    \item If the results are not up to expectation, perform optimization over the $\str$ hyperparameter using standard AutoML methods like Bayesian Optimization (BO). This could be automated especially if a fidelity metric is provided, but manual sequential checks could also be used given just 1 hyperparameter and a query-efficient approach like BO.
\end{enumerate}

In practice, if \algname is widely adopted, an open research or developer community would also likely be able to share such configurations and fine-tuning, similar to how fine-tuned deep learning models are also being shared today. Even if \algname is implemented by closed-source companies, economies of scale would make it worth fine-tuning and optimizing \algname across languages and text types.

\subsection{Refinement of watermarked text to improve fidelity} \label{app:refine}
As paraphrasing is applied to the original text when performing the watermark, there might be a change in the style of writing, some loss in information, or in the case of code watermarking, loss of functionality. However, these can be mitigated through several techniques, some of which we have already implemented in our experiments.

In practice, the client could assess the fidelity of the watermarked text $T_w$ before using it. If $T_w$ does not meet the fidelity threshold (i.e., semantic content is not sufficiently preserved), the client could simply use the LLM paraphraser to correct the watermarked text $T_w$ to increase semantic preservation. This could be done automatically as demonstrated in the code example (e.g., Reflexion, or multiple generations), or done manually with prompt engineering. The LLM paraphraser will once again introduce the same embedded watermark to produce the new watermarked text $T_w'$, strengthening both the verifiability and fidelity of the text.

Additionally, as the field develops, it is expected for LLMs' paraphrasing capabilities to increase significantly across domains, languages and text types. This enables the \algname framework, using these more capable LLMs, to generate watermarked text with smaller and smaller semantic degradation, further improving its performance and allowing \algname to remain effective in highly specialized or technical domains.

\subsection{Additional implementation methods like Beam search}

Instead of performing conventional sampling of the perturbed logits during the watermarking text generation phase, we could apply beam search for generation. This allows us to generate several alternative versions of the watermarked text, while ensuring that the tokens being selected retains fidelity and achieves verifiability by selecting high scoring logits with strong watermark perturbations. This provides a more computationally efficient alternative to the multi-round refinement in \cref{app:refine}.
Finally, the final watermarked text can be selected from the list of generations based on some criterion that balances fidelity and verifiability.
We have implemented a version of this in our code release, available at \url{https://github.com/aoi3142/Waterfall}.

\end{document}